\documentclass[10pt]{iopart}
\pdfoutput=1
\usepackage{fullpage}
\usepackage{iopams}
\usepackage{setstack}
\usepackage{braket}
\usepackage{ifthen}
\usepackage{tikz}
\usepackage[pdftex,breaklinks=true,bookmarks=false,colorlinks=true,citecolor=violet,urlcolor=blue,linkcolor=black]{hyperref}
\usepackage{caption}
\usepackage{subcaption}

\newcommand{\journ}[5]
{\ifthenelse{\equal{#1}{jsm}}{
{\it #5}, #4 J. Stat. Mech. \href{http://iopscience.iop.org/1742-5468/#4/#2/#3}{#3}}
{\ifthenelse{\equal{#1}{pr}}{
{\it #5}, #4 Phys. Rev {\bf #2} \href{http://link.aps.org/abstract/PR/v#2/e#3}{#3}}
{\ifthenelse{\equal{#1}{prl}}{
{\it #5}, #4 Phys. Rev. Lett {\bf #2} \href{http://link.aps.org/abstract/PRL/v#2/e#3}{#3}}
{\ifthenelse{\equal{#1}{prb}}{
{\it #5}, #4 Phys. Rev. B {\bf #2} \href{http://link.aps.org/abstract/PRB/v#2/e#3}{#3}}
{\ifthenelse{\equal{#1}{pra}}{
{\it #5}, #4 Phys. Rev. A {\bf #2} \href{http://link.aps.org/abstract/PRA/v#2/e#3}{#3}}
{\ifthenelse{\equal{#1}{arxiv}}{
{\it #5}, #4 \href{http://arxiv.org/abs/#2.#3}{arXiv:#2.#3}}
{\ifthenelse{\equal{#1}{rmp}}{
{\it #5}, #4 Rev. Mod. Phys {\bf #2} \href{http://link.aps.org/abstract/RMP/v#2/e#3}{#3}}
{\ifthenelse{\equal{#1}{cond-mat}}{preprint
\href{http://arxiv.org/abs/cond-mat/#2}{cond-mat/#2}}
{\ifthenelse{\equal{#1}{pre}}{
{\it #5}, #4 Phys. Rev. E {\bf #2} \href{http://link.aps.org/abstract/PRE/v#2/e#3}{#3}}
{\it #5}, #4 #1 {\bf #2} #3}}}}}}}}
}

\newcommand{\journdoi}[6]{{\it #5}, #4 #1 {\bf #2} \href{http://dx.doi.org/#6}{#3}}

\newcommand{\journlink}[6]{{\it #5}, #4 #1 {\bf #2} \href{#6}{#3}}

\newcommand{\journnolink}[5]{{\it #5}, #4 #1 {\bf #2} #3}

\begin{document}

\title{Emptiness formation probability, Toeplitz determinants, and conformal field theory} 

\author{Jean-Marie St\'ephan}
\address{Physics Department, University of Virginia, Charlottesville, VA 22904-4714}
\eads{\mailto{jean-marie.stephan@virginia.edu}}
\begin{abstract}

We revisit the study of the emptiness formation probability, the probability of forming a sequence of $\ell$ spins with the same ferromagnetic orientation in the ground-state of a quantum spin chain. We focus on two different examples, exhibiting strikingly different behavior: the XXZ and Ising chains. One has a conserved number of particles, the other does not. In the latter we show that the sequence of fixed spins can be viewed as an additional boundary in imaginary time. We then use conformal field theory (CFT) techniques to derive all universal terms in its scaling, and provide checks in free fermionic systems. These are based on numerical simulations or, when possible, mathematical results on the asymptotic behavior of Toeplitz and Toeplitz+Hankel determinants. A perturbed CFT analysis uncovers an interesting $\ell^{-1}\log \ell$ correction, that also appears in the closely related spin full counting statistics. The XXZ case turns out to be more challenging, as scale invariance is broken. We use a simple qualitative picture in which the ferromagnetic sequence of spins freezes all degrees of freedom inside of a certain ``arctic'' region, that we determine numerically. We also provide numerical evidence for the existence of universal logarithmic terms, generated by the massless field theory living outside of the arctic region. 

\end{abstract}
\maketitle

 \tableofcontents
 \vfill\eject

\hypersetup{linkcolor=red}
\section{Introduction}
  \subsection{Emptiness formation probability in spin chains}

In recent years, a great deal of effort has been put into the understanding and calculation of correlations in integrable one-dimensional quantum critical systems. Although many physical properties can be understood from the Bethe-ansatz solution \cite{Bethe,Gaudin}, the explicit calculation of physical correlation functions is a much more formidable task. Much progress has however occurred in this direction, in particular in the spin-1/2 XXZ spin chain, a key physical example where such methods are successful. New methods have been developed to compute such correlations and their asymptotics exactly, starting from the lattice model  \cite{Quantuminverse,Jimbo1,Jimbo2,Lyon1,Lyon2}. Moreover, the use of conformal field theory (CFT) \cite{BPZ} provides a way to access various universal exponents in a simple and transparent way. However, it does not give access to non-universal quantities, and usually still requires some input from more exact methods \cite{BIK}. These two complementary approaches have also been 
shown to agree in several studies, see e.g. \cite{Lyon3,Lyon4}.

In the first framework, arguably the simplest quantity one can compute is the probability of forming a sequence of $\ell$ consecutive spins with the same up orientation, the so-called emptiness formation probability (EFP). For this reason it has already been subject to quite extensive studies  \cite{EFP_first,EFP_first2,EFP_XX,EFP_XXZrazumov,EFP_XXZrazumov2,AbanovKorepin,KorepinLukyanov,EFP_Kozlowski,EFP_Cantini}, especially (but not always) in the simpler limit of an infinite chain. In particular, many exact results have been derived, expressing the EFP using multiple integral or determinantal representations, from which an asymptotic expansion can sometimes be extracted. Note that the study of the EFP is also interesting in -- simpler -- free fermionic systems. Indeed in this case the EFP usually boils down to a Toeplitz determinant, whose interesting mathematical properties have attracted much attention since the celebrated Kaufman-Onsager spontaneous magnetization problem \cite{Ising_exponent,
FisherHartwig}. 

Two intriguing features stand out in the various asymptotic results obtained for the EFP. First, the leading term can depend on the model. 
 The logarithm of the EFP (in short $\log EFP$) is proportional to $\ell^2$ for the XXZ chain \cite{EFP_XX,EFP_XXZrazumov}, while it only scales proportionally to $\ell$ for the Ising chain \cite{FranchiniAbanov}. This behavior can be traced back to the fact that in the XXZ chain magnetization is conserved, which is not the case for Ising: in fermionic language the particle number is conserved in one case, but \emph{not} in the other. Second, the EFP is sensitive to criticality. When the system is critical, power law corrections -- logarithmic terms for the logarithmic EFP -- to the leading term start to appear. While these terms can often be explained at the purely mathematical level (e.g. singularities in the symbol generating the Toeplitz determinant), they have so far remained unexplained at the field-theoretical level. \\

This paper is a partial attempt at filling this gap. We show that the Ising case can be understood in full detail using CFT. As a consequence, we confirm the universality of the power law correction derived by Franchini and Abanov \cite{FranchiniAbanov}, and show that it is a simple fraction of the central charge of the CFT. We provide several generalizations to open systems and finite geometries, where we compute a universal scaling function, in the spirit of entanglement entropy studies. All these results are extensively checked in free-fermionic systems, combining numerical evaluations and asymptotic results on Toeplitz and Toeplitz+Hankel determinants. Following \cite{SD_2013}, we then carry out a perturbed CFT analysis of the subleading corrections, and show how the first goes slightly beyond the known Toeplitz results. As we shall see it takes a form proportional to $\ell^{-1}\log \ell$, with a coefficient determined by the central charge, the geometry of the system, and a single non universal short 
distance cutoff, the so-called extrapolation length \cite{Lebowitz}. We also study several related quantities, 
including the R\'enyi-Shannon entropy and full counting statistics, where this type of subleading correction makes an appearance as well. \textcolor{black}{All these results also apply to more complicated CFTs with non conserved number of particles, such as the minimal models. In particular this provides with a method to identify the central charge.}

In contrast, the XXZ case turns out to be much more subtle. We are nevertheless able to gather some evidence for the universality of the power law corrections, and show how the general scaling may be explained in terms of an ``arctic phenomenon'', in which all degrees of freedom are frozen in a region of area proportional to $\ell^2$ in imaginary time. Such behavior is familiar in the study of domino tilings on the Aztec diamond. In this picture 
a massless field theory, whose exact properties are unknown, lives outside of a ``frozen region'' that we determine numerically in various free fermionic examples. 
\pagebreak
  \subsection{Problem studied}
Let us consider the following quantum Hamiltonian, so called XY chain in transverse field.
  \begin{equation}\label{eq:xychain}
 H=-\sum_{j=1}^L\left[\left(\frac{1+\nu}{2}\right)\sigma_j^x\sigma_{j+1}^x+\left(\frac{1-\nu}{2}\right)\sigma_j^y\sigma_{j+1}^y+h\sigma_j^z\right].
\end{equation}
We will study this Hamiltonian with periodic boundary conditions ($\sigma_{L+1}^\alpha=\sigma_1^\alpha$, $\alpha=x,y$), as well as open boundary conditions
 ($\sigma_{L+1}^\alpha=0$, $\alpha=x,y$). The system is in the ground-state $|\psi\rangle$, and the spins are measured in the basis of the eigenstates of the $\sigma_j^z$. What makes this model particularly simple is that using a Jordan-Wigner transformation, it can be rewritten in terms of free fermions \cite{LiebSchultzMattis}:
 \begin{equation}
  H=-\sum_{j=1}^{L}\left(c_i^\dag c_{i+1}+\nu\,c_{i}^\dag c_{i+1}^\dag +h.c\right)-h\sum_{j=1}^L\left(2c_j^\dag c_j-1\right),
 \end{equation}
where $c_{L+1}^\dag=e^{i\pi \hat{N}}c_1^\dag$ for periodic boundary conditions, and $c_{L+1}^\dag=0$ for open boundary conditions. Here $\hat{N}=\sum_{j=1}^{L} c_j^\dag c_j$ is the fermion number operator. The dictionary between spins and fermions is fairly simple: a fermion at site $j$ corresponds to an up spin, while a vacant site $j$ corresponds to a down spin.
 
 As is well known, this model has several critical lines, as a function of the parameters $\nu$ and $h$. The first we consider is $\nu>0$ and $h=1$, and belongs to the Ising universality class\footnote{We ignore here the other similar line $\nu>0,h=-1$, because it involves changes in conformal boundary conditions. We discuss this in Sec.~\ref{sec:EFP_down}.}. Its low-energy properties are described by the Majorana fermion, a CFT with central charge $c=1/2$. In particular, the point $h=1,\nu=1$ is the Ising chain in transverse field (ICTF). The other ($\nu=0, |h|<1$) has a vanishing fermion pairing term, and therefore an additional $U(1)$ symmetry. It belongs to the free compactified boson universality class (or Luttinger liquid), and has central charge $c=1$. In the following, we will mostly focus on these two critical 
lines.
 \begin{figure}[ht]
\begin{subfigure}[b]{0.5\textwidth}
\centering
\includegraphics{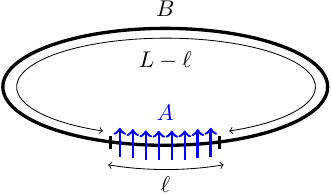}
 \caption{Periodic chain}
 \label{fig:efp_periodic}
\end{subfigure}
\begin{subfigure}[b]{0.5\textwidth}
\centering
\includegraphics{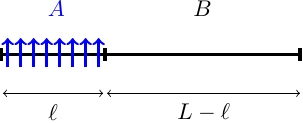}
 \caption{Open chain}
 \label{fig:efp_open}
\end{subfigure}
\caption{Emptiness formation probability for a sequence of $\ell=8$ consecutive up spins in a system of length $L$. For the open chain the fixed spins are the $\ell$ first.}
\label{fig:efp}
\end{figure}\\
One may then ask what is the probability of observing a sequence of $\ell$ consecutive up spins, somewhere in the chain (in the open chain, the sequence ends on the left edge). More precisely, let us denote by $A$ the subsystem formed by the $\ell$ spins, and $B$ the rest. See Fig.~\ref{fig:efp} for a picture. The reduced density matrix for $A$ is
\begin{equation}
 \rho_A={\rm Tr}_B |\psi\rangle\langle \psi|,
\end{equation}
and the EFP is defined as
\begin{equation}
 \mathcal{P}=\langle \uparrow\uparrow\ldots\uparrow|\rho_A|\uparrow\uparrow\ldots\uparrow\rangle.
\end{equation}
For later convenience, we also introduce the logarithmic emptiness formation probability ($\log EFP$)
\begin{equation}
 \mathcal{E}=-\log \mathcal{P}.
\end{equation}
We wish to study the asymptotic behavior of $\mathcal{E}$ with the size $\ell$, while keeping the aspect ratio fixed to a certain fixed value $\ell/L \in [0;1)$. We are mainly interested in identifying universal terms in the process. In the following we use the notation $\mathcal{E}_p$ in the periodic system, and $\mathcal{E}_o$ in the open system. 
  \subsection{Outline and summary of the results}
  The paper is organized as follows. In Sec.~\ref{sec:cbc} we study the example of the Ising critical line, and show that the ferromagnetic string acts as an additional boundary in euclidean time. We then use boundary CFT methods to compute universal terms in the $\log EFP$. In particular we obtain
 \begin{eqnarray}\label{eq:main1}
 \mathcal{E}_p&=&a_1\ell+\frac{c}{8}\log \left[\frac{L}{\pi}\sin\left(\frac{\pi \ell}{L}\right)\right]+
 a_0^{(p)}-\frac{\xi c}{8}\cot \left(\frac{\pi\ell}{L}\right)\frac{\log \ell}{L}+\mathcal{O}\left(\frac{1}{\ell}\right),\\\label{eq:main2}
 \mathcal{E}_o&=&a_1\ell-\frac{c}{16}\log\left[\frac{4L}{\pi}\frac{\tan^2\left(\frac{\pi \ell}{2L}\right)}{\sin \left(\frac{\pi\ell}{L}\right)}\right]
 +a_0^{(o)}+\frac{\xi c}{32}\times \frac{2-\cos \left(\frac{\pi \ell}{L}\right)}{\sin\left(\frac{\pi \ell}{L}\right)} \frac{\log \ell}{L}+\mathcal{O}\left(\frac{1}{\ell}\right).
\end{eqnarray} 
$c$ is the (universal) central charge. $a_1,a_0^{(p)},a_0^{(o)}$ are non universal, we compute them using asymptotic results on Toeplitz and Toeplitz + Hankel determinants in Sec.~\ref{sec:toeplitz}. 
The ``extrapolation length'' $\xi$ is also a non universal quantity, known in the context of surface critical phenomena \cite{Lebowitz}. 
We show that
\begin{equation}
 \xi=\frac{1}{2\nu}
\end{equation}
for the Ising critical line of the XY chain. 
The $L^{-1}\log \ell$ terms can be seen as semi-universal \cite{SD_2013}: once $\xi$ is fixed they are solely determined by the central charge of the CFT, and the geometry of the system. 
For the EFP this type of correction goes slightly beyond the available mathematical results on Toeplitz determinants, and offers an interesting generalization to a recently raised conjecture \cite{Kozlowski_conjecture}. 
\textcolor{black}{Our main result (\ref{eq:main1}) also applies to any unitary CFT with central charge $c<1$. The generalization of the open case (\ref{eq:main2}) to any such CFTs is also straightforward.}

 We then turn our attention (Sec.~\ref{sec:noncbc}) to the XXZ chain, as an example of a free $c=1$ boson CFT, and an important exception. This is known to exhibit strikingly different behavior, with in particular the $\log EFP$ scaling as
 \begin{equation}\label{eq:xxzscaling}
  \mathcal{E}_{p,o}=a_2^{(p,o)}\ell^2+a_1^{(p,o)}\ell+b_0^{(p,o)} \log \ell+\mathcal{O}(1).
 \end{equation}
We propose a simple qualitative picture for such a behavior, in which the ferromagnetic string in imaginary time generates a region of area $\propto \ell^2$ where the degrees of freedom are frozen. Crucial to this argument is the fact that magnetization is conserved, which is not the case for Ising. A massless field theory, which is \emph{not} a CFT, then lives at the exterior of this region. This picture is a manifestation of the well-known arctic phenomenon for dimers. We also provide numerical evidence for the universality of $b_0$, by studying different lattice realizations and varying the aspect ratio $\ell/L$. The analytical derivation of $b_0$ using field-theoretical methods is however left as an important open problem.

Finally in Sec.~\ref{sec:other}, we briefly discuss some related problems to which our results could be applied, including the Shannon entropy and full counting statistics.
\section{The ferroelectric string as a conformal boundary condition}
\label{sec:cbc}
This section mainly relies on a simple observation: the configuration with $\ell$ up spins, viewed in euclidean time, should renormalize to a conformal boundary condition \cite{Cardy1989}. 
We explain why this is in Sec~\ref{sec:imag}, before applying it to the calculation of universal logarithmic terms (Sec~\ref{sec:logarithmic_terms}) and semi-universal corrections (Sec~\ref{sec:loglsl_corr}).
\subsection{Imaginary time picture}
\label{sec:imag}
\begin{figure}[htbp]
  \begin{subfigure}[b]{0.5\textwidth}
\centering
 \includegraphics{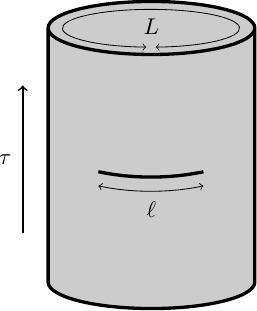}
 \caption{Infinite cylinder with a slit}
 \label{fig:euclidean_cylinder}
 \end{subfigure}
 \begin{subfigure}[b]{0.5\textwidth}
  \centering
 \includegraphics{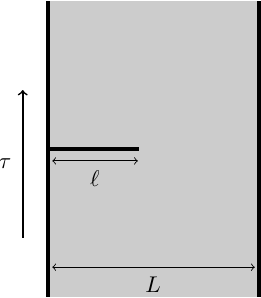}  
\caption{Infinite strip with a slit}
\label{fig:euclidean_strip}
 \end{subfigure}
\caption{Space-time geometry (the time direction is vertical). In the cylinder geometry, there are two corners with angle $2\pi$. In the strip geometry only one corner has angle $2\pi$, but there are two additional corners with angle $\pi/2$.}
\label{fig:euclidean}
\end{figure}

\textcolor{black}{
The projector onto the ground-state $\ket{\psi}\bra{\psi}$ of the Hamiltonian $H$ can formally be seen as the result of an infinite imaginary time evolution
\begin{equation}
 e^{-\tau H}\underset{\tau \to \infty}{\sim} e^{-\tau E_0}\ket{\psi}\bra{\psi}
\end{equation}
where $E_0$ is the ground-state energy. Using this, the EFP can be rewritten as
\begin{equation}\label{eq:imag}
 \mathcal{P}=\lim_{\tau \to \infty}\frac{\bra{s}e^{-\tau H}\delta(\ket{\sigma}-\ket{\uparrow\ldots\uparrow})e^{-\tau H}\ket{s}}{\braket{s|e^{-2\tau H}|s}},
\end{equation}
where $\ket{s}$ is any state not orthogonal to the ground-state, and $\delta(\ket{\sigma}-\ket{\uparrow\ldots\uparrow})$ selects the configuration with all spins up in the $z$-basis. 
However the classical spins in the 2d Ising model correspond, before taking the Hamiltonian limit, to the eigenstates of the $\sigma_j^x$. Therefore, a configuration with all spins up in the $z$-basis translates into
\begin{eqnarray}
 |\!\uparrow\uparrow\ldots\uparrow\rangle_z&=&\frac{1}{2^{\ell/2}}\, \left(|\!\rightarrow\rangle_x+|\!\leftarrow\rangle_x\right)
  \left(|\!\rightarrow\rangle_x+|\!\leftarrow\rangle_x\right)
  \ldots  \left(|\!\rightarrow\rangle_x+|\!\leftarrow\rangle_x\right)\\
  &=&\frac{1}{2^{\ell/2}}\sum_{\{\sigma_j^x=\rightarrow,\leftarrow\}} |\sigma_1^x \sigma_2^x\ldots \sigma_\ell^x\rangle\\
  &=&|{\rm free}\rangle_x,
\end{eqnarray}
a state where all orientations are allowed. From the point of view of the 2d Ising model, this corresponds to imposing a free boundary condition on the spins. This lattice free boundary condition is
 expected to renormalize to a \emph{free conformal boundary condition} \cite{Cardy1989}.
}
Hence, taking (minus) the logarithm of (\ref{eq:imag}) yields
 \begin{equation}\label{eq:lefp_z}
 \mathcal{E}_p=-\log\left(\frac{\mathcal{Z}_{\rm cyl}^{(\rm slit)}}{\mathcal{Z}_{\rm cyl}}\right).
\end{equation}
$\mathcal{Z}_{\rm cyl}$ is the partition function of an infinite cylinder, and $\mathcal{Z}_{\rm cyl}^{(\rm slit)}$ is the partition function of an infinite cylinder with a slit inserted in the middle, as is shown in Fig.~\ref{fig:euclidean_cylinder}. In the open geometry the cylinders are replaced by strips, see Fig.~\ref{fig:euclidean_strip}. Since the external boundary conditions are also free, there are no changes in boundary conditions. In both cases, the logarithmic EFP is given by the corresponding differences in free energies $F=-\log \mathcal{Z}$. 
 \\
From general arguments, each free energy has to leading order a ``bulk'' contribution proportional to the area $L\times 2\tau$ of the cylinder (strip), but this contribution is cancelled in the ratio (\ref{eq:lefp_z}) defining the $\log EFP$. Then, there is a ``line'' contribution proportional to the boundary lengths, which are $2L+\ell$ for the slitted cylinder (strip), and $2 L$ for the cylinder (strip). Hence the leading contribution to the $\log EFP$ will be proportional to $\ell$. More precisely, we expect
\begin{equation}\label{eq:general_scaling}
\mathcal{E}=-\log \mathcal{P}=a_1\ell+b_0\log L+f(\ell/L)+a_0+o(\ell^0).
\end{equation}
$a_1$ is the lineic free energy contribution of the slit. Let us now explain the next two terms. As was first pointed out by Cardy and Peschel \cite{CardyPeschel}, the logarithmic divergence is a general consequence of the presence of corners in the partition functions we have to evaluate. Indeed, there are two corners with angle $2\pi$ in $\mathcal{Z}_{\rm cyl}^{(\rm slit)}$, and one $2\pi$ as well as two $\pi/2$ corners in $\mathcal{Z}_{\rm strip}^{(\rm slit)}$. The coefficient $b$ takes a simple form: for each corner with angle $\theta$ there is a contribution \cite{CardyPeschel}
\begin{equation}\label{eq:CardyPeschel}
 \Delta F=\left[\frac{c}{24}\left(\frac{\theta}{\pi}-\frac{\pi}{\theta}\right)+\frac{\pi}{\theta}h_{bcc}\right]\log L
\end{equation}
to the free energy. $c$ is the central charge of the corresponding conformal field theory, with $c=1/2$ for the Ising universality class. $h_{bcc}$ is the dimension of a possible boundary changing operator, but there are none here ($h_{bcc}=0$) since all boundaries are free. We therefore have $b_0=c/8=1/16$ in the periodic geometry, and $b_0=-c/16=-1/32$ in the open geometry. \textcolor{black}{The non-trivial function $f(\ell/L)$ of the aspect ratio $\ell/L$ is there because of the long-range correlations in our critical system. We calculate it in the next section using CFT.} $a_0$ is a constant. Notice that both $a_1$ and $a_0$ are cutoff-dependent and therefore non-universal. We calculate them for the Ising chain in Sec.~\ref{sec:toeplitz}, using exact results on the asymptotics of Toeplitz and Toeplitz + Hankel determinants. 

Let us finally mention that the coefficient $b_0$ has a discontinuity for the aspect ratio $\ell/L=1$: in this limit the corners disappear in the periodic geometry, and become four $\pi/2$ corners in the open geometry. We refer to \cite{Stephan_Ising,Zaletel} for a study of the corresponding universal terms. In the following, we shall focus solely on the case $0\leq \ell/L<1$.
  \subsection{Logarithmic terms}
  \label{sec:logarithmic_terms}
      \paragraph{CFT derivation.---}
  The universal scaling function $f(\ell/L)$ can be obtained as follows. We use complex coordinates $w=x+i\tau$ on the cylinder ($x \in [0,L)$). 
  Let $\mathcal{C}$ be a contour around the slit, and $w_0=x_0+i\tau_0$ a point on the slit. We consider the infinitesimal transformation
  $w\mapsto w+\delta \ell\Theta(x-x_0)$ inside the contour, and $w\mapsto w$ outside. \textcolor{black}{$\Theta$ is the Heaviside step function}. Such a transformation stretches the slit by $\delta \ell$ in the horizontal direction, while keeping $L$  unchanged. The variation of the free energy is encoded in the $T_{xx}$ component of the stress-energy tensor as
  \begin{equation}
   \delta F=\frac{\delta \ell}{2\pi}\int_{{\cal C}}\langle T_{xx}\rangle \,dw.
  \end{equation}
In complex coordinates we have $T_{xx}=T(w)+\bar{T}(\bar{w})$. The integral can be obtained by mapping back the slitted cylinder to the upper half plane $\mathbb{H}=\{z\in \mathbb{C},{\rm Im}\,z>0\}$, where $\langle T(z)\rangle_{\mathbb{H}}=0$ due to translational invariance. The transformation law of the stress tensor is
\begin{equation}
 T(z)=\left(w^\prime(z)\right)^2T(w)+\frac{c}{12}\{w(z),z\},
\end{equation}
where $\{w(z),z\}$ denotes the Schwarzian derivative of $w(z)$
\begin{equation}
 \{w(z),z\}=\frac{w^{\prime\prime\prime}(z)}{w^\prime(z)}-\frac{3}{2}\left(\frac{w^{\prime\prime}(z)}{w^{\prime}(z)}\right)^2.
\end{equation}
The contributions of the holomorphic and anti-holomorphic parts of the stress-tensor are identical, and we get
\begin{equation}\label{eq:someint}
 \frac{\delta F}{\delta \ell}=-\frac{c}{12\pi}\int_{w^{-1}(\mathcal{C})}S(z)\,dz\qquad,\qquad S(z)= \left(w^{\prime}(z)\right)^{-1}\{w(z),z\}.
\end{equation}
The conformal transformation from the upper-half plane is given by\footnote{A possible way to find this transformation is to map the cylinder to the plane, using $u=e^{2i\pi w/L}$. The image of the slit is a circular arc with radius $1$. This arc can be mapped to $\mathbb{R}^+$ using a Moebius transformation $\zeta=(u-u_1)/(u-u_2)$, where $u_1$ and $u_2$ are the images of the two endpoints of the slit. The resulting geometry is sent to the upper-half plane using $z=\sqrt{\zeta}$. This finally gives us $z(w)$, which may be inverted to get $w(z)$.} 
  \begin{equation}
 w(z)=\frac{L}{2i\pi} \log \left[\frac{1+z^2e^{i\pi \ell/L}}{z^2+e^{i\pi \ell/L}}\right]
\end{equation}
and the integral (\ref{eq:someint}) can be evaluated using the Residue theorem. We find
\begin{equation}
 \frac{\delta F}{\delta \ell}=\frac{\pi c}{8L\tan (\pi \ell/L)}.
\end{equation}
This can be integrated to $F=\int_\epsilon^\ell \delta F$. $\epsilon$ is an UV cutoff of the order of a lattice spacing, which gives back the contribution from the Cardy-Peschel formula (\ref{eq:CardyPeschel}). The other contribution is
\begin{equation}
 f(\ell/L)=\frac{c}{8}\log \left(\sin\frac{\pi \ell}{L}\right),
\end{equation}
and we finally get
\begin{equation}\label{eq:log_periodic}
 \mathcal{E}_p-a_1\ell=\frac{c}{8}\log \left[\frac{L}{\pi} \sin \left(\frac{\pi \ell}{L}\right)\right]+a_0^{(p)}+o(1),
\end{equation}
where $a_0^{(p)}$ is a non universal constant. Notice that such a ``chord-length'' scaling is quite common, as it appears for example in the study of the entanglement entropy \cite{CalabreseCardy2004}, albeit with a different prefactor. Also in our case this is a subleading term, due to the presence of a line free energy.

\begin{figure}[htbp]
 \begin{subfigure}[b]{0.5\textwidth}
   \includegraphics[width=7.5cm]{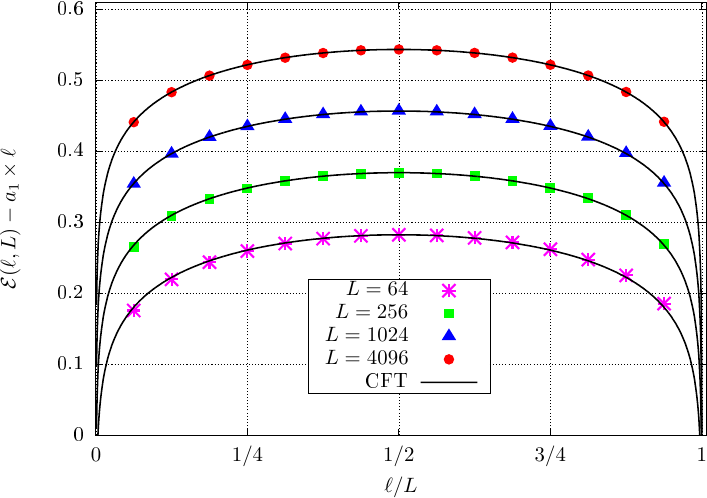}
 \caption{Periodic geometry}
 \label{fig:log_periodic}
 \end{subfigure}
\begin{subfigure}[b]{0.5\textwidth}
  \includegraphics[width=7.5cm]{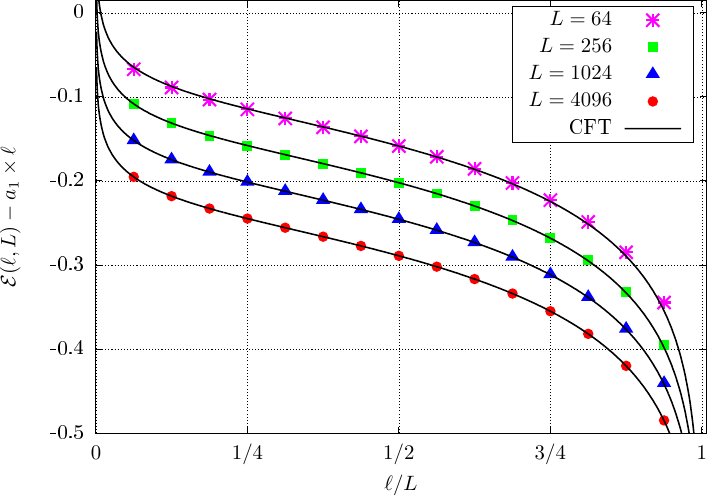}
 \caption{Open geometry}
 \label{fig:log_open}
\end{subfigure}
 \caption{Numerical extraction of the logarithmic terms (Eqs.~(\ref{eq:log_periodic}, \ref{eq:log_open})). We show the data for system sizes $L=64,256,1024,4096$ and the aspect ratios $\ell/L=1/16,2/16,\ldots,15/16$. The regular spacings between the curves reflect the global scaling of the EFP, $(c/8)\log L$ for (a) and $-(c/16)\log L$ for (b).}
 \label{fig:log}
\end{figure}

The open geometry can be treated as well. The conformal transformation from the upper half-plane is modified to\footnote{The strip is mapped to the upper half-plane using $u=e^{i\pi w/L}$. The image of the slit is again a circular arc with radius $1$, which is mapped to a slit $[0;i]$ using a Moebius transformation $\zeta(u)$, and mapped to the upper half plane using $z=\sqrt{1+\zeta^2}$.}
\begin{equation}
 w(z)=\frac{2L}{\pi}\arctan \left[\tan\left(\frac{\pi \ell}{2L}\right)\sqrt{1-z^2}\right],
\end{equation}
and we find
\begin{equation}\label{eq:log_open}
 \mathcal{E}_p-a_1\ell=-\frac{c}{16}\log\left[\frac{4L}{\pi}\frac{\tan^2\left(\frac{\pi \ell}{2L}\right)}{\sin \left(\frac{\pi\ell}{L}\right)}\right]+a_0^{(o)}+o(1)
\end{equation}
after a similar calculation. It is important that the $\ell$ up spins stand at the beginning of the open chain. If this is not the case the slit does not touch the left boundary any more, and the corresponding space-time geometry has the topology of an annulus, as opposed to the upper-half plane in our derivation. Global conformal invariance would then not be sufficiently strong to constrain the finite-size scaling function $f(\ell/L)$, and this would make its determination much more difficult.
    \paragraph{Numerical checks.---}
We provide here some numerical evidence to support our main results (\ref{eq:log_periodic}) and (\ref{eq:log_open}), focusing on the Ising chain in transverse field. As is well known, such a chain can be studied by a mapping onto free fermions \cite{LiebSchultzMattis}. The calculation of the EFP essentially boils down to a determinant, as follows from the Wick theorem. See e.g Refs. \cite{FranchiniAbanov,Stephan_Ising,Rectangle1} for the details. In particular we have
\begin{eqnarray}\label{eq:detp}
 \mathcal{P}_p=\det_{1\leq i,j\leq \ell}\left(\frac{\delta_{ij}}{2}+\frac{1}{2L}\csc\left[\frac{\pi (i-j+1/2)}{L}\right]\right),
\end{eqnarray}
in the periodic case, and 
\begin{equation}\label{eq:deto}
 \mathcal{P}_o=\det_{1\leq i,j\leq \ell}\left(\frac{\delta_{ij}}{2}+\frac{1}{4L+2}\csc\left[\frac{\pi (i-j+1/2)}{2L+1}\right]+\frac{1}{4L+2}\csc\left[\frac{\pi (i+j-1/2)}{2L+1}\right]\right)
\end{equation}
in the open case. Here $\csc$ denotes the cosecant function
\begin{equation}
\csc x=\frac{1}{\sin x}.
\end{equation}
It is then straightforward to compute these determinant numerically, using standard Linear Algebra routines. The results are shown in Fig.~\ref{fig:log}, where we plot
$\mathcal{E}-a_1\ell$ for system sizes up to $L=4096$ and several aspect ratios $\ell/L$. \textcolor{black}{For $a_1$ we used the exact results in section.~\ref{sec:toeplitz}, which gives $a_1=\log 2-\frac{2C}{\pi}$, where $C$ is Catalan's number. It can also be obtained with very good accuracy by fitting $\mathcal{E}$ to Eq.~(\ref{eq:main1},\ref{eq:main2}) for any aspect ratio $\ell/L$.}
As can be seen from the plot, the numerics agree very well with the CFT result, and even more so for the largest system sizes. Notice that in the periodic geometry, the invariance under $\ell\to L-\ell$ is only exact in the asymptotic $L\to \infty$ limit. Indeed, we shall see in the next subsection that the first subleading correction is actually \emph{antisymmetric} under $\ell\to L-\ell$, and both these prediction are non-trivial. Let us mention that these results can be checked away from the ICTF ($h=1$, $\nu\neq 1$), as is shown in Sec~\ref{sec:universality}.
  \subsection{An unusual subleading correction}
  \label{sec:loglsl_corr}
Let us now discuss lattice effects, and the possible subleading corrections they generate. In general these may be understood by adding perturbations by local operators to the CFT action. The operators can live in the bulk, or be localized near the boundary. In any case they need to be consistent with the symmetries of the physical system, and also be irrelevant (or at worst, marginal). Indeed, any relevant operator would drive the system to a different bulk CFT, or a different conformal boundary condition.

Here we focus on a generic perturbation by the stress-tensor itself along the boundary, following \cite{SD_2013}. The stress-tensor is arguably always one of the leading irrelevant operators, and we shall see that it actually produces the leading finite-size effects in our problem. Such a perturbation also admits an interesting physical interpretation \cite{DubailReadRezayi}. Let us modify the CFT action to
\begin{equation}\label{eq:perturb}
 S\to S+\frac{\xi}{2\pi}\int_{\rm slit}dx\, T_{xx},
\end{equation}
 where the perturbation is localized near the slit. The coupling $\xi$ has the dimension of a length, and is typically of the order of one lattice spacing. Such a perturbation effectively pushes the conformal boundary to a distance $\xi$, to make up for the fact that the physical (lattice) boundary and the (RG invariant) conformal boundary have no reason to lie at the exact same position. For this reason, $\xi$ is nothing but the ``extrapolation length'', a standard concept in boundary critical phenomena \cite{Lebowitz}. Let us also mention that this perturbation plays a key role in the study of quantum quenches \cite{CalabreseCardy2007} as well as the entanglement spectrum in fractional quantum Hall states \cite{DubailReadRezayi}.
    \paragraph{CFT derivation.---}
Here we reproduce the main arguments of Ref.~\cite{SD_2013}, and refer to it for the detail. In the presence of a corner with angle $2\pi$, perturbing by the stress tensor produces to leading order an interesting correction of the form $L^{-1}\log \ell$, whose prefactor may be exactly determined for simple enough geometries. In general the contribution of the perturbation (\ref{eq:perturb}) to the free energy is given by the following power series
\begin{equation}\label{eq:general_correction}
 \Delta F=-\sum_{n=1}^\infty \left(\frac{-\xi}{2\pi}\right)^n\frac{1}{n!}\int_{\rm slit} dx_1 \int_{\rm slit} dx_2\;\ldots \int_{\rm slit} dx_n \,\langle T_{xx}(x_1)T_{xx}(x_2)\ldots T_{xx}(x_n)\rangle.
\end{equation}
Let us explain the physical origin of the logarithmic correction on the simplest example of the infinite geometry. In this limit ($L\to\infty$), the conformal transformation simplifies into
\begin{equation}
 w(z)=\frac{\ell}{2}\frac{z^2-1}{z^2+1}.
\end{equation}
The slit is the interval $[-\ell/2,\ell/2]$ and its image through the inverse conformal mapping is $\mathbb{R}_+$. Using $T_{xx}=T(w)+\overline{T}(\overline{w})$:
\begin{equation}\label{eq:var_freenrj}
 \Delta F=-\frac{\xi c}{\pi}\int_{-\ell/2+\epsilon}^{\ell/2-\epsilon} \langle T(w)\rangle\,dw=\frac{\xi c}{12\pi} \int_{w^{-1}(-\ell/2+\epsilon)}^{w^{-1}(\ell/2-\epsilon)} S(z)\,dz.
\end{equation}
We have used once again the mapping to the upper-half plane and $\langle T(z)\rangle=0$. The regulator $\epsilon$ in Eq.~(\ref{eq:var_freenrj}) is of the order of a lattice spacing, and acts as an ultraviolet cutoff. $S(z)$ is given here by
\begin{equation}
 S(z)=\frac{3}{2\ell}\left(\frac{1}{2z^3}-\frac{1}{z}+\frac{z}{2}\right)
\end{equation}
For $\ell$ large we also have $w^{-1}(-\ell/2+\epsilon)\sim \sqrt{\epsilon/\ell}$ and $w^{-1}(\ell/2-\epsilon)\sim\sqrt{\ell/\epsilon}$, and this allows to get the subleading contributions to the free energy. The term proportional to $z$ in $S(z)$ generates a correction $\sim 1/\epsilon$ of order one, and the term proportional to $1/z^3$ generates a correction in $\sim \epsilon \ell^{-2}$. The most interesting contribution comes from the $1/z$ term
\begin{equation}\label{eq:simple_loglsl}
 \Delta \tilde{F}=-\frac{\xi c}{8\pi} \frac{\log (\ell/\epsilon)}{\ell},
\end{equation}
which is the $\ell^{-1}\log \ell$ advertised earlier. Notice that (\ref{eq:simple_loglsl}) also yields a contribution proportional to  $\ell^{-1}\log \epsilon$. This term is much less interesting to us, but it is important to take it into account for numerical purposes (see below). The $\ell^{-1}\log \ell$ is specific to corners with angle $2\pi$. It can be shown that for an angle $\theta$, the corresponding contribution to $S(z)$ takes the form $z^{1-\theta/\pi}$, which can only produce a logarithm after integration when $\theta=2\pi$.

\begin{figure}[htbp]
 \begin{subfigure}[b]{0.5\textwidth}
   \includegraphics[width=7.3cm]{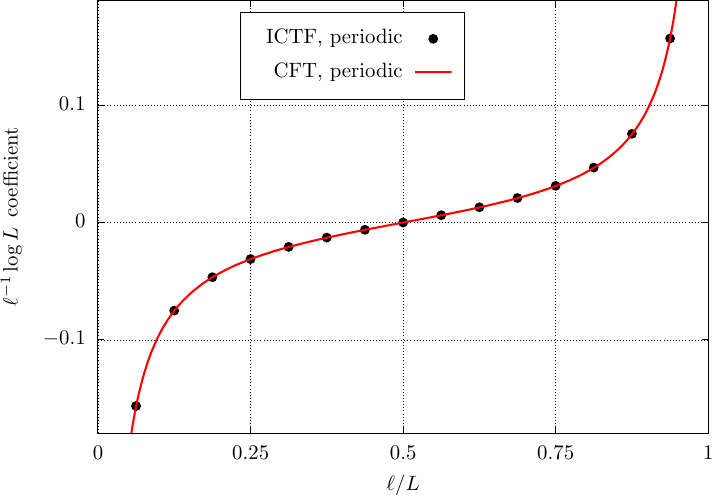}
 \caption{Periodic geometry}
 \label{fig:loglsl_periodic}
 \end{subfigure}
\begin{subfigure}[b]{0.5\textwidth}
  \includegraphics[width=7.3cm]{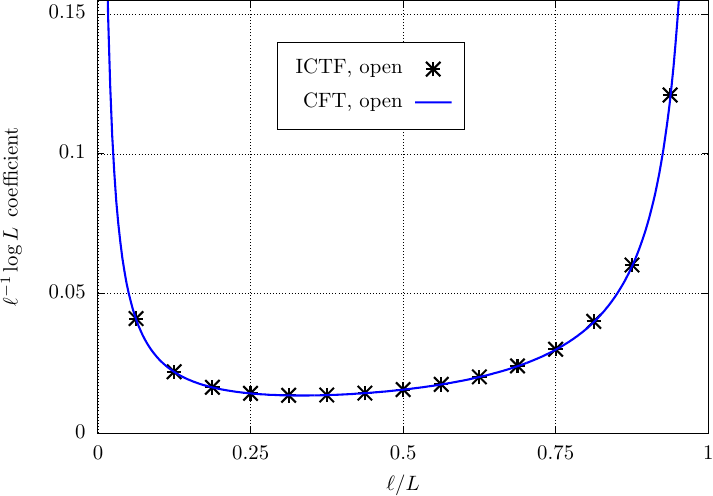}
 \caption{Open geometry}
 \label{fig:loglsl_open}
\end{subfigure}
 \caption{Numerical extraction of the $L^{-1}\log \ell$ term for the ICTF in both geometries, for aspect ratios $\ell/L=1/16,2/16,\ldots, 15/16$. The agreement with the CFT predictions of Eqs.~(\ref{eq:loglsl_periodic}, \ref{eq:loglsl_open}) is close to perfect, with a relative error smaller than to one percent.}
 \label{fig:loglsl}
\end{figure}

In the more general finite-size case we use the result of \cite{SD_2013}. The contribution of interest is given by 
\begin{equation}
g(\ell/L)=\frac{\xi c}{24\pi}\left(\sum_c\frac{w^{\prime\prime}(z_c)}{|w^{\prime\prime}(z_c)|}{\rm Res}\left[S(z);z=z_c\right]\right)\times \log \ell
\quad,\quad S(z)=\left(w^\prime(z)\right)^{-1}\{w(z),z\},
\end{equation}
where the sum runs over all the $2\pi$ corners, their positions being $w(z_c)$. $w(z)$ is the conformal transformation from the upper half plane. The open geometry has only one such corner, and applying this formula we obtain
\begin{equation}\label{eq:loglsl_open}
 g_{o}(\ell/L)=\frac{\xi c}{32}\times \frac{2-\cos \left(\frac{\pi \ell}{L}\right)}{\sin\left(\frac{\pi \ell}{L}\right)}\times \frac{\log \ell}{L}.
\end{equation}
A similar calculation, with two corners, yields
\begin{equation}\label{eq:loglsl_periodic}
 g_p(\ell/L)=-\frac{\xi c}{8}\times \cot \left(\frac{\pi \ell}{L}\right)\times \frac{\log \ell}{L}
\end{equation}
in the periodic geometry. This term reproduces Eq.~(\ref{eq:simple_loglsl}) in the limit $\ell/L\to 0$. Eqs.~(\ref{eq:loglsl_periodic}, \ref{eq:loglsl_open}) can be seen as ``semi-universal''. While they are proportional to the extrapolation length which is a non universal quantity, the central charge also appears and the shape as a function of $\ell/L$ is purely geometric: it does not depend on the specifics of the model. Another way to see this is to replace the dominant length scales $L$ and $\ell$ by $L+\epsilon$ and $\ell+\varepsilon$, where $\epsilon$ and $\varepsilon$ are short distance cutoffs, of the order of a lattice spacing. Performing this substitution in Eq.~(\ref{eq:general_scaling}) and expanding again, it is easy to see that the $L^{-1}\log \ell$ terms remain unaffected. 
    \paragraph{Numerical checks.---}
The extrapolation length is a so far unknown function of $\nu$ on the Ising critical line. It should a priori depend on the boundary condition, but it has been found that $\xi=1/2$ for free boundary conditions in the ICTF \cite{SD_2013}. which we study here.
 To make contact with our predictions we consider the combination
 \begin{equation}
\mathcal{E}_p-a_1\ell-\frac{c}{8}\log \left[\frac{L}{\pi} \sin \left(\frac{\pi \ell}{L}\right)\right]
 \end{equation}
 in the periodic geometry, and a similar combination in the open case. We then fit this to $d_0+d_1L^{-1}\log \ell+d_2 L^{-1}$ for various system sizes up to $L=4096$ and several aspect ratios $\ell/L$. The results are summarized in Fig.~\ref{fig:loglsl}, and show again a very good agreement with our prediction. We emphasize the importance of adding a last term $d_2L^{-1}$ to remove the otherwise large finite-size effects. 
\subsection{Universality of the emptiness formation probability}
\label{sec:universality}
\begin{figure}[htbp]
\begin{subfigure}[b]{0.5\textwidth}
   \includegraphics[width=7.3cm]{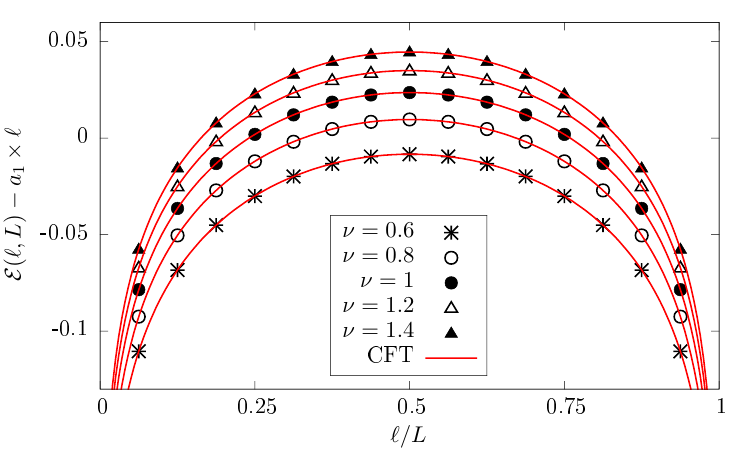}
 \caption{Logarithmic terms (\ref{eq:log_periodic})}
 \label{fig:XY_log}
 \end{subfigure}
\begin{subfigure}[b]{0.5\textwidth}
  \includegraphics[width=7.3cm]{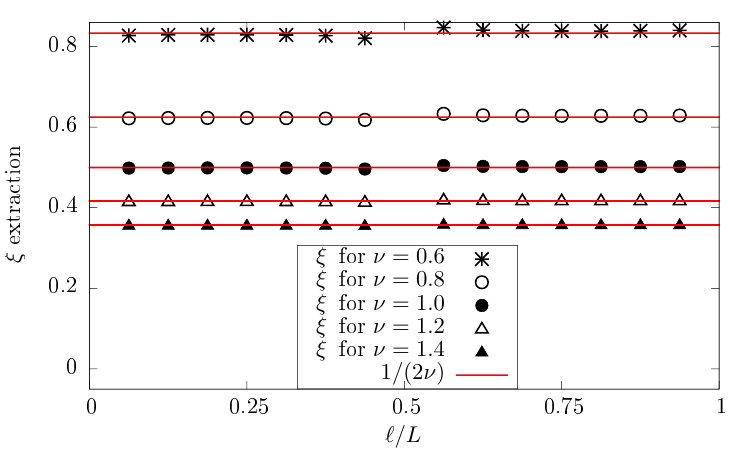}
 \caption{Extrapolation length extraction}
 \label{fig:XY_loglsl}
\end{subfigure}
\caption{(a): Logarithmic terms $\mathcal{E}(\ell,L)-a_1\ell$ for system size $L=4096$ and anisotropies $\nu=0.6,0.8,1,1.2,1.4$. The shift between the curves is due to the non-universal constant $a_0^{(p)}$. (b): Extraction of the extrapolation length $\xi(\nu)$, by plotting the ratio of Eq.~(\ref{eq:ratio}) for the same values of $\nu$ as in (a). Red curves are our conjecture $\xi(\nu)=1/(2\nu)$, which reproduces the data very well. We use the system sizes $L=3072,3584,4096$ for the fit.}
\label{fig:XY}
 \end{figure}

We discuss in this section the universality of the results we derived. Let us first focus on the $XY$ chain in transverse field for general $\nu>0$ (only $\nu=1$ was considered before). We consider a periodic system, but the open case is similar. The $EFP$ is given by
\begin{equation}
 \mathcal{P}_{p}=\det_{1\leq i,j\leq \ell}\left(\frac{\delta_{ij}}{2}+\frac{1}{2L}\sum_{q}\cos\left[q(i-j)+\theta_q\right]\right)
\end{equation}
where the sum runs over $q=(2m+1)\pi/L$, $m=-L/2,\ldots L/2-1$, and $\theta_q=\arctan(\nu \tan \frac{q}{2})$. Its numerical computation for large system sizes remains straightforward.  
Since the XY chain belongs to the Ising universality class for any $\nu>0$, the logarithmic terms derived in section~\ref{sec:logarithmic_terms} are expected to still hold. This is indeed the case, as is shown in Fig.~\ref{fig:XY_log}. We plot here the leading logarithmic term $\mathcal{E}(\ell,L)-a_1\ell$ for a very large total system size and various values of $\nu$. The agreement with CFT remains perfect.

The $\ell^{-1}\log \ell$ correction turns out to be more interesting, as the perturbed CFT result (\ref{eq:loglsl_periodic}) is proportional to the extrapolation length $\xi$, which depends on the details of the lattice model and therefore on $\nu$. We extract $\xi(\nu)$ in Fig.~\ref{fig:XY_loglsl}, by plotting the ratio
\begin{equation}\label{eq:ratio}
 \frac{L^{-1}\log \ell {\rm \;\;term} }{-\frac{c}{8}\cot \frac{\pi \ell}{L}}
\end{equation}
for various aspect ratios $\ell/L$ and using the same procedure as at the end of Sec~\ref{sec:loglsl_corr} ($\nu=1$). Provided Eq.~(\ref{eq:loglsl_periodic}) is correct, this should converge to $\xi(\nu)$, independent on the aspect ratio. As can be seen the numerical results agree very well with this prediction. From them we are also able to conjecture the simple formula
\begin{equation}\label{eq:xi_conj}
 \xi(\nu)=\frac{1}{2\nu}
\end{equation}
for the extrapolation length. This is compatible with the known value $\xi(\nu=1)=1/2$ for the ICTF, and will be computed in the next section.

The Ising CFT is the simplest example of an unitary CFT with central charge $c<1$, the so-called unitary minimal models (or their subsets). All these CFT have a finite number of primary fields, and so a finite number of allowed conformal boundary conditions \cite{Cardy1989}. 
These minimal models describe the long-distance physics of several critical spin chains. For example, one can study a generalization of the Ising chain corresponding to the Hamiltonian limit of the $Q=3$ states Potts model \cite{QPotts}, or anyonic chains \cite{Goldenchain} related to RSOS models \cite{AndrewsBaxterForrester} and their generalizations \cite{Vincent1,Vincent2}.

For all these models one can define an analogue of the configuration with all up spins, and this homogeneous configuration is expected to renormalize to one the allowed conformal boundary conditions. This determination has been done for many CFTs (see e.g. \cite{SB,Cardy1989,AOS}). For a periodic system the derivation of the leading logarithmic term is not sensitive to the boundary condition, provided it is conformal. Hence we expect the result
\begin{equation}\label{eq:perbcc}
 \mathcal{E}_{p}-a_1\ell=\frac{c}{8}\log \left[\frac{L}{\pi}\sin \left(\frac{\pi\ell}{L}\right)\right]+o(1)
\end{equation}
to hold for all these models. In an open system one needs to be more careful, as the conformal boundary condition on the slit can happen to differ from the external boundary condition (see Fig.~\ref{fig:euclidean_strip}). In this case one needs to introduce boundary changing operators, and a generalization of the calculation presented in Sec~\ref{sec:logarithmic_terms} yields
\begin{equation}\label{eq:openbcc}
 \mathcal{E}_o-a_1\ell=\frac{c}{16}\log \left[\frac{L}{\pi}\sin \left(\frac{\pi \ell}{L}\right)\right]+\left(4h-\frac{c}{8}\right)\log \left[\frac{L}{\pi}\tan\left(\frac{\pi \ell}{2L}\right)\right]+o(1).
\end{equation}
$h$ is the dimension of the boundary changing operator \cite{Cardy1989}. Such a phenomenon can also occur in the Ising chain, if one looks at the probability of having all spins down, see Sec.~\ref{sec:EFP_down}. We also expect the $L^{-1}\log \ell$ to be present, however for these models there is no guarantee that this will be the leading correction, and this can make its numerical extraction much more difficult in practice.

An important exception is the XXZ chain, where the configuration with all up spins is not expected to renormalize to a conformal boundary condition. This is due to two facts: (i) the number of particles is conserved by the XXZ Hamiltonian, and (ii) the sequence of up spins injects a lot of particles in the system. For all the models that share this property we expect (\ref{eq:perbcc},\ref{eq:openbcc}) to break down. However, our boundary CFT arguments should still apply if an alternating sequence $\ket{\uparrow\downarrow\uparrow\downarrow\ldots\uparrow\downarrow}$ is imposed instead of $\ket{\uparrow\uparrow\uparrow\uparrow\ldots\uparrow\uparrow}$. Indeed contrary to the latter, the former does not inject any particles in the system. This is discussed in Sec~\ref{sec:mutualinf}. For all models that share the two (i) and (ii) properties, the boundary is not conformal anymore. This is discussed in Section~\ref{sec:noncbc}.
\section{Relation to the theory of Toeplitz determinants}
\label{sec:toeplitz}
We now turn our attention to the limit $L/\ell\to\infty$, so that $\ell$ becomes the only remaining length scale in the problem. 
 The results of the previous section simplify into
 \begin{eqnarray}\label{eq:expansion_p}
  \mathcal{E}_p&=&a_1\ell+\frac{c}{8}\log \ell+a_0^{(p)}-\frac{\xi c}{8\pi}\frac{\log\ell}{\ell}+\mathcal{O}(1/\ell),\\\label{eq:expansion_o}
  \mathcal{E}_o&=&a_1\ell-\frac{c}{16}\log \ell+a_0^{(o)}+\frac{\xi c}{32\pi}\frac{\log\ell}{\ell}+\mathcal{O}(1/\ell).
 \end{eqnarray}
For the Ising critical line $c=1/2$. $\xi$ is a so far analytically unknown function of $\nu$ on this line, but $\xi(\nu=1)=1/2$ for the ICTF. Interestingly, the determinants of Eqs.~(\ref{eq:detp}) and (\ref{eq:deto}) for the ICTF now take an even simpler form:
\begin{eqnarray}\label{eq:det_inf_p}
 \mathcal{P}_{p}&=&\det_{0\leq i,j\leq \ell-1}\left(g_{i-j}\right),\\\label{eq:det_inf_o}
  \mathcal{P}_{o}&=&\det_{0\leq i,j\leq \ell-1}\left(g_{i-j}+g_{i+j+1}\right),
\end{eqnarray}
with
\begin{equation}\label{eq:matelements}
 g_k=\frac{\delta_{k0}}{2}+\frac{(-1)^k}{2\pi(k+1/2)}.
\end{equation}
The factor $(-1)^k$ doesn't change the determinant, but has been added to make contact with the existing literature. There are slightly more complicated formulas for general $\nu$. 
 An important feature of these determinants is that the matrix elements do not depend\footnote{Contrary to the determinants (\ref{eq:detp}, \ref{eq:deto}), because of the fixed non-zero $\ell/L$ aspect ratio.} on the size $\ell$, and this allows for a more rigorous treatment, using known mathematical results. 
In the following we summarize some of these, regarding the asymptotics of Toeplitz (\ref{eq:det_inf_p}) and Toeplitz+Hankel (\ref{eq:det_inf_o}) determinants.
 This presentation is heuristic, and does not make any attempt at mathematical rigor. We refer to Refs.~\cite{Toeplitz_r0,Toeplitz_r1,Toeplitz_r2} for reviews. We then use these results to find the exact asymptotic expansion of the determinants corresponding to Eq.~(\ref{eq:matelements}).  
 Let us finally mention that the first three terms (but not the fourth) in Eq.~(\ref{eq:expansion_p}) have already been derived by Franchini and Abanov \cite{FranchiniAbanov}; we will come back to this point later. 
\subsection{The strong Szeg\"o limit theorem and the Fisher-Hartwig conjecture}
It is customary to see the matrix elements as Fourier coefficients of a certain generating function, usually called \emph{symbol}:
\begin{eqnarray}
 g_k=[g]_k=\frac{1}{2\pi}\int_{0}^{2\pi}g(\phi)e^{-ik\phi}\qquad,\qquad g(\phi)=\sum_{k\in \mathbb{Z}}\,[g]_k\, e^{ik\phi}.
\end{eqnarray}
The following central result is due to Szeg\"o \cite{Szegostrong}. If $g(\phi)$ is a sufficiently smooth, non vanishing function with winding number $0$, the asymptotic expansion of $\det T_\ell(g)=\det_{0\leq i,j\leq \ell-1} \left([g]_{i-j}\right)$ when $\ell\to \infty$ is given by
\begin{equation}
\det (T_\ell(g))\sim \left(G[g]\right)^\ell E[g],
\end{equation}
where $G[g]$ is the geometric mean of $g(\phi)$
\begin{equation}
  G[g]=\exp\left([\log g]_0\right),
\end{equation}
and $E[g]$ is given by
\begin{equation}
 E[g]=\exp\left(\sum_{k=1}^{\infty}k[\log g]_k[\log g]_{-k}\right).
\end{equation}
The leading exponential term can be intuitively understood by noticing that $T_\ell$ is almost invariant with respect to translations. Writing down $T_\ell$ in Fourier space yields for large $\ell$
\begin{equation}
 U_\ell^\dag T_\ell U_\ell\approx {\rm diag}\left[g(0),g(2\pi/\ell),\ldots,g(2\pi(\ell-1)/\ell)\right]\qquad,\qquad (U_\ell)_{jl}=e^{-i2\pi jl/\ell},
\end{equation}
provided the $[g]_k$ decay sufficiently fast. The determinant is then dominated by the geometric mean to the $\ell-$th power. Interestingly this result is exponentially accurate. 
There are however many interesting physical problems for which the $[g]_k$ decay slowly, and the symbol has singularities. This especially tends to happen when studying critical systems. Among them one can mention the famous Ising magnetization correlator at the critical point \cite{Ising_exponent}, the monomer correlator for dimers on the square lattice \cite{FisherStephenson}, full counting statistics \cite{Abanov_stats1,Abanov_stats2}, entanglement in critical 1d spin chains \cite{Korepin1,Korepin2,Korepin3}, just to name a few. The following result \cite{BasorToeplitz,Bottcher,EhrhardtPhD,BasorEhrhardt} will be useful for us. Consider generating functions that only possess pointwise ``phase'' discontinuities, parametrized  as
\begin{equation}\label{eq:fh_form}
 g(\phi)=f(\phi) \prod_{r=1}^R\exp\left(i\beta_r\arg\left[e^{i\left(\phi-\phi_r\right)}\right]\right),
\end{equation}
with $-\pi<\phi_1<\phi_2<\ldots<\phi_R\leq \pi$, $|\beta_r|<1/2$ and $f(\phi)$ sufficiently smooth, e.g. satisfying the hypothesis of the strong Sz\"ego limit theorem. We use the convention $\arg z\in (-\pi;\pi]$. It is also possible to account for root-type singularities of the form $(2-2\cos \phi)^{\alpha}$, but we discard them here to make the discussion simpler. The asymptotic formula for $\det_{0\leq i,j\leq \ell-1} \left([g]_{i-j}\right)$ is modified to
\begin{equation}\label{eq:fh}
\det (T_\ell(g))\sim \left(G[f]\right)^\ell \ell^{\omega}\tilde{E}[g]\qquad,\qquad \omega=-\sum_{r=1}^{R}\beta_r^2
\end{equation}
where $\tilde{E}[g]$ is a somewhat complicated constant. The most interesting physical piece in such formulae are the power law correction $\ell^\omega$, because they usually correspond to universal physical terms.

Such a prescription is known as the Fisher-Hartwig conjecture \cite{FisherHartwig}, and many instances of this conjecture and various generalizations are by now theorems \cite{BasorTracy,DeiftItsKrasovsky}. In the following we will also need similar results for the closely related Toeplitz+Hankel determinants of the form $\det((T+H)_\ell)=\det_{0\leq i,j\leq \ell-1}\left([g]_{i-j}+[g]_{i+j+1}\right)$.
\subsection{Rigorous asymptotic expansion of the EFP}
For the Ising critical line of the XY chain the symbol reads \cite{FranchiniAbanov}
\begin{equation}\label{eq:gen_symbol}
 g(\phi)=\frac{1}{2}+\frac{1}{2}\exp\left(-i\arctan\left[\nu \tan \frac{\phi}{2}\right]\right),
\end{equation}
\textcolor{black}{It has an obvious phase discontinuity, due to the divergence of $\tan \frac{\phi}{2}$ at $\phi=\pi$. It can be parametrized as in Eq.~(\ref{eq:fh_form}), with $R=1$, $\phi_1=0$ and $\beta_1=\beta=-1/4$. $f(\phi)$ is given by
\begin{eqnarray}
 f(\phi)=|g(\phi)|=\frac{1}{2}\sqrt{2+2\cos\left(\arctan\left[\nu \tan \frac{\phi}{2}\right]\right)}
\end{eqnarray}}
and has winding number zero, as it should. Armed with this, it is now straightforward to apply the results of \cite{BasorEhrhardt}, and get the leading terms in the asymptotic expansion of the emptiness formation probability. 
\paragraph{Periodic case.---}
We recover here the result of \cite{FranchiniAbanov}, which reads
\begin{equation}
 \mathcal{E}=a_1\ell+\frac{1}{16}\log \ell+a_0^{(p)}+o(1),
\end{equation}
compatible with the CFT prediction. Note that the coefficient of the logarithm is solely determined by $\beta_1$, which does not depend on $\nu$. This is also expected from CFT (see. Eq.~\ref{eq:expansion_p}).
The non universal constants $a_1$ and $a_0^{(p)}$ depends on $\nu$, and are given by
\begin{eqnarray}
a_1&=&[\log f]_0\\
 a_0^{(p)}&=&-\sum_{k=1}^{\infty}k ([\log f]_k)^2-\log (G(3/4)G(5/4))
\end{eqnarray}
$G$ is the Barnes G-function \cite{Barnes}, solution of the functional relation $G(z+1)=\Gamma(z)G(z)$, where $\Gamma$ is the Euler Gamma function. 
For example at $\nu=1$, we get the exact $a_1=\log 2-2\frac{C}{\pi}$, as well as $a_0^{(p)}=0.0951599254(1)$ by truncating the sum to the first $5000$ terms in the sum. $C$ is the Catalan constant.
\paragraph{Open case.---}
We apply here the result of Ref.~\cite{BasorEhrhardt}, and refer to it for the general formula. Let us just mention that the critical power law exponent is modified to $\omega=-\frac{\beta}{2}-\frac{3}{2}\beta^2=\frac{1}{32}$ in this particular case. Once again this exponent does not depend on $\nu$, and we find
\begin{equation}
 \mathcal{E}=a_1\ell-\frac{1}{32}\log \ell+a_0^{(o)}+o(1),
\end{equation}
in agreement with CFT (Eq.~\ref{eq:expansion_o}). $a_1$ is identical to the periodic case, and the other non universal term $a_0^{(o)}$ is given by
\begin{eqnarray}
 a_0^{(o)}=-\frac{1}{2}\sum_{k=1}^{\infty}k ([\log f]_k)^2-\frac{C}{4\pi}
 -\log\left[\frac{G(3/4)^2G(5/4)}{G(1/2)2^{7/32}\pi^{1/8}}\right]
\end{eqnarray}
\textcolor{black}{Simarly to the periodic case, we get e.g. $a_0^{(o)}=-0.02149124775(5)$ at $\nu=1$.} 
\subsection{Possible subleading corrections}
\label{sec:toeplitz_subleading}
Let us now discuss possible sources of subleading corrections to the general Fisher-Hartwig formula. If one relaxes the condition on the $\beta_r$ in Eq.~(\ref{eq:fh_form}), the parametrization is not necessarily unique anymore, because each $\beta_r$ can be shifted by arbitrary integers $\beta_r\to\beta_r+n_r$, the zero winding condition on $f$ being ensured by $\sum n_r=0$. The following result, often dubbed ``generalized Fisher-Hartwig'' has been conjectured by Basor and Tracy \cite{BasorTracy} and proved recently by Deift, Its and Krasovsky \cite{DeiftItsKrasovsky}:
\begin{equation}\label{eq:gfh}
 \det(T_\ell(g))\sim\left(G[f]\right)^\ell\sum_{\{n_r\}}^\prime \ell^{\omega(\{\beta_r\},\{n_r\})}  E[g;\{n_r\}],
 \quad,\quad \omega(\{\beta_r\},\{n_r\})=-\sum_{r=1}^{R}(\beta_r+n_r)^2
\end{equation}
where the sum runs over all possible $n_r\in \mathbb{Z}$ subject to the condition $\sum_{r=1}^{R}n_r=0$. It is important to stress that the symbol $\sim$ stands for equivalent. This formula therefore only differs from Eq.~(\ref{eq:fh}) when several different representations have the same exponent ${\rm Re}\,w(\{\beta_r\},\{n_r\})$.

It is however tempting to assume that all the terms in (\ref{eq:gfh}) appear as subleading corrections to the determinant. 
Several studies \cite{CalabreseEssler,FagottiCalabrese,Gutman,Abanov_stats1,Abanov_stats2} supported this claim, by showing that at least the first subleading branches can be identified in the asymptotic expansion. Moreover, each branch can separately be improved by adding corrections in the form of a power series in $\ell^{-1}$. Based on explicit computations of the first subleading terms, Kozlowski conjectured \cite{Kozlowski_conjecture} that this prescription yields, provided $f(\phi)$ is smooth, a full asymptotic expansion of the determinant. More precisely, it is given by
\begin{equation}\label{eq:kozlowski_conjecture}
 \det(T_\ell(g))=\left(G[f]\right)^\ell\sum_{\{n_r\}}^\prime \ell^{\omega(\{\beta_r\},\{n_r\})}  E[g;\{n_r\}] \left(1+\sum_{i=1}^{\infty} \alpha_{\{\beta_r\},\{n_r\}}^{(i)}\ell^{-i}\right).
\end{equation}
In our case the generating function has only one Fisher-Hartwig singularity, which means there is no other subleading representation\footnote{We point out a slight inaccuracy in Ref.~\cite{FranchiniAbanov}, where the authors considered a  parametrization with $\beta_1+1$ and conjectured a subleading term in $\ell^{-(\beta_1+1)^2+\beta_1^2}=\ell^{-1/2}$ in the $\log EFP$. Such a parametrization is however forbidden, because $f(\phi)$ doesn't have winding number zero in this case.}. Because of the cusp at $\phi=\pi$, $f(\phi)$ is continuous but not smooth, hence our $\ell^{-1}\log \ell$ does not contradict Eq.~(\ref{eq:kozlowski_conjecture}). 
The Riemann-Hilbert analysis of Ref.~\cite{Kozlowski_conjecture} can however be generalized to account for less regular symbols, by introducing the parametrization
\begin{equation}\label{eq:new_parametrization}
 g(\phi)=f(\phi)\left(1+z\right)^{-\mu(z)}\left(1+1/z\right)^{-\bar{\mu}(z)}\quad,\quad z=e^{i\phi},
\end{equation}
where $\mu$ and $\bar{\mu}$ are analytic functions of $z$ in some neighborhood of the unit circle, and chosen in such a way that $f(\phi)$ be also analytic\footnote{I am grateful to Karol Kozlowski for explaining this to me.}. For symbols $g(\phi)$ that can be written in such a way, the Riemann-Hilbert approach allows to obtain the form of the subleading corrections: in general each coefficient $\alpha_{\{\beta_r\},\{n_r\}}^{(i)}$ in Eq.~(\ref{eq:kozlowski_conjecture}) should likely be replaced by a polynomial of degree at most $i$ in $\log \ell$. One can check that writing
\begin{eqnarray}
 \mu(z)&=&-\beta-\sum_{p=1}^{\infty}\eta_p (z+1)^p\\
 \bar{\mu}(z)&=&\beta-\sum_{p=1}^{\infty}\eta_p (1/z+1)^p,
\end{eqnarray}
with $\beta=-1/4$ as before preserves the phase discontinuity at $\theta=\pi$. Notice that $\eta_p=0$ for all $p$ would give back the previous parametrization of Eq.~(\ref{eq:fh_form}). The $\eta_p$ have to be set so as to smoothen the more regular part of the symbol $f(\phi)$ around the singularity at $\theta=\pi$. Let us determine the first few for the Ising critical line of the XY chain, where the symbol is given by Eq.~(\ref{eq:gen_symbol}). 
$\eta_1=-\frac{1}{4\pi\nu}$ is obtained by imposing $f^\prime(\pi)=0$, and $\eta_2=\eta_1/2$ is the only value that makes the second derivative $f^{\prime\prime}(\pi)$ finite. It turns out the knowledge of $\eta_1$ is sufficient for our purpose. Pushing further the analysis of Ref.~\cite{Kozlowski_conjecture} we get a term of the form
\textcolor{black}{
\begin{equation}
 2\beta^2\eta_1\times \frac{\log \ell}{\ell}=-\frac{1}{32\pi \nu}\times \frac{\log \ell}{\ell}
\end{equation}
to the $\log EFP$. Comparison with the CFT prediction (\ref{eq:expansion_p}) yields an extrapolation length 
\begin{equation}
\xi(\nu)=\frac{1}{2\nu},
\end{equation}
consistent with our previous numerical results of Sec.~(\ref{sec:loglsl_corr}), as well as the result $\xi(\nu=1)=1/2$ of \cite{SD_2013}.} We also study a generalization, relevant to the full counting statistics, in Sec.~\ref{sec:fcs}. Finally, let us mention that a similar structure of logarithmic corrections has already been studied for closely related Fredholm determinants of the generalized sine kernel (see e.g. Refs.~\cite{Sinekernel1,Sinekernel2}).

\begin{figure}[ht]
\centering 
\includegraphics[width=9cm]{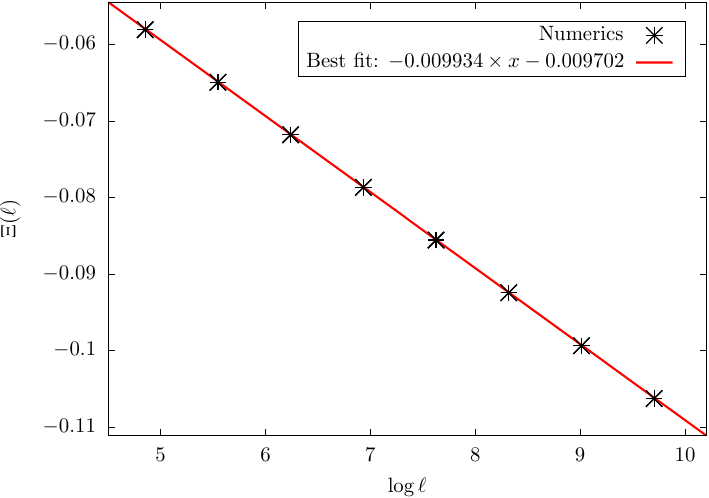}
 \caption{$\Xi(\ell)$ for system sizes $\ell=128,256,512,1024,2048,4096,8192,16384$. The data is plotted as a function of $\log \ell$ and agrees very well with the scaling predicted from CFT. }
 \label{fig:efp_inf_num}
\end{figure}

Since our results are based on heuristic arguments, we are unable to prove the existence of the $\ell^{-1}\log \ell$ term. To numerically resolve this issue, it is convenient to form the combination
\begin{equation}
 \Xi(\ell)=\left(\mathcal{E}-a_1\ell-b_0\log \ell-a_0\right)\times \ell
\end{equation}
either in the periodic or open case. Provided the CFT result is correct, this quantity should scale as
\begin{equation}\label{eq:xi_scaling}
 \Xi(\ell)=\gamma \log \ell+\delta +o(1).
\end{equation}
Fig.~\ref{fig:efp_inf_num} shows some numerical data for the ICTF in the periodic case, where $\Xi(\ell)$ is plotted against $\log \ell$, for various system sizes between $\ell=128$ and $\ell=16384$. The data follows an almost straight line, and confirms the scaling of Eq.~(\ref{eq:xi_scaling}). Our best numerical estimate is $\gamma=-0.00994(2)$ which agrees with the predicted $\gamma=-1/(32\pi)$ within less than 0.1 percents of relative error. 
We also observed that this estimate can be improved by adding further subleading corrections. We found that adding two terms of the form $b_{-2}\ell^{-2}\log^2 \ell$ and $a_{-2}\ell^{-2}$ reproduces the numerical data very well, and brings the relative error on $\gamma$ under $4.10^{-6}$.
A similar (but not identical) scaling has also been observed \cite{StephanPhD,SD_2013} for the fidelity considered in \cite{DubailStephan_2011}, where a closed-form formula can be derived and a systematic asymptotic expansion be performed. In this case there is a $\ell^{-1}\log \ell$ as well, but the subleading corrections take the somewhat simpler form of $\ell^{-p}\log \ell$ and $\ell^{-p}$ terms, with $p\geq 1$. 

Let us now summarize our findings. In general we expect the $\log EFP$ to scale as
\begin{equation}\label{eq:general_expansion}
 \mathcal{E}=a_1 \ell+b_0\log \ell+a_0+\sum_{p=1}^{\infty} \left(\sum_{k=0}^{p} a_{p,k}\left(\log \ell\right)^{k}\right)\times \ell^{-p}
\end{equation}
where the prefactor of the $\ell^{-p}$ is a polynomial of degree at most $p$ in $\log \ell$. Such a scaling is natural in terms of higher-order corrections to the Toeplitz determinant, but also in CFT. Indeed, taking into account the other contributions of the perturbed expansion (\ref{eq:general_correction}) leads to a similar structure for the corrections, although the precise determination of all terms becomes very quickly cumbersome. Note however that not all logarithmic terms in the expansion (\ref{eq:general_expansion}) are semi-universal, in the sense defined in Sec.~\ref{sec:loglsl_corr}: making the substitution $\ell\to \ell+\epsilon$ in Eq.~(\ref{eq:general_expansion}) and expanding again, we observe that a $(\log \ell)^{k}\ell^{-p}$ remains unchanged only if all the lesser subleading corrections are of the form $(\log \ell)^{k'}\ell^{-p'}$, with $k'<k$ and $p'<p$. In particular this implies that the $(\log \ell)^k \ell^{-k}$ are automatically semi-universal, but the other terms usually not; the former would be the most interesting to study within CFT.   
\section{The ferroelectric string as a boundary that breaks scale-invariance}
\label{sec:noncbc}
We now turn our attention to the XXZ chain, and related models with conserved magnetization. The Hamiltonian of the XXZ chain reads
  \begin{eqnarray}
   H&=&\sum_{j} \left(\sigma_j^x\sigma_{j+1}^x+\sigma_j^y\sigma_{j+1}^y+\Delta \sigma_j^z\sigma_{j+1}^z\right).
  \end{eqnarray}
It is well established that the low energy physics of such systems is described by a Luttinger liquid CFT\cite{LL} in the range $-1< \Delta\leq 1$. In imaginary time this is a Gaussian model, with action
\begin{equation}\label{eq:action}
 S=\frac{\kappa}{4\pi}\int \left(\nabla \varphi\right)^2 dxdy\qquad,\qquad \varphi\equiv \varphi+2\pi r
\end{equation}
The ``height'' field $\varphi$ lives on a circle of radius $r$. 
The physical parameter that governs the decay of the correlations is the compactification radius $R=r\sqrt{2\kappa}$. For the XXZ chain it is given by
\begin{equation}\label{eq:cradius}
 R(\Delta)=\sqrt{2-\frac{2}{\pi}\arccos \Delta}.
\end{equation}
In the following we will explain the scaling of the $\log EFP$ for general $\Delta$ using simple imaginary time arguments. Roughly speaking, the XXZ Hamiltonian conserves the particle number (or magnetization), and imposing a long sequence of up spins injects a lot of particles in the system. The slit of Sec.~\ref{sec:cbc} then becomes a non trivial region and this breaks conformal invariance. 
We will also check this argument at the free-fermion points $\Delta=0$ (or $R=1$), where numerical computations of these regions simplify considerably.
\subsection{Scaling of the EFP}
Many exact results are available for the scaling of the EFP, both for the infinite and semi-infinite geometries. Generically we have
  \begin{equation}
   \mathcal{E}=a_2 \ell^2+a_1\ell+b_0\log \ell+o(1),
  \end{equation}
so that the leading term is proportional to $\ell^2$, in contradction with and the analysis of Sec.~\ref{sec:cbc} for Ising and minimal models. We explain the precise physical origin of this leading term in the next subsection. Intuitively the difference with Ising comes from the fact that the ferroelectric string injects a lot of particles  (or up spins) in a system where the total number of particles is fixed (or that has zero total magnetization). We also want to provide numerical evidence for the universality of the subleading logarithms in \ref{sec:logterms}. To try and achieve both goals it is convenient to introduce another free fermionic spin chain with radius $R=1$:
\begin{equation}\label{eq:dimer_chain}
 H_d=\sum_j \left[c_{2j}^\dag\left(c_{2j-2}+c_{2j-1}+3c_{2j}+c_{2j+1}+c_{2j+2}\right)+c_{2j+1}^\dag\left(c_{2j}+c_{2j+1}+c_{2j+2}\right)\right]+h.c
\end{equation}
Just like the $XXZ$ chain has the same eigenvectors as the transfer matrix of the six-vertex model, $H_d$ has the same eigenvectors as the transfer matrix of the dimer model on the square lattice \cite{LiebTM,Ikhlef,Stephan09}. Hence we dub (\ref{eq:dimer_chain}) the ``dimer chain''. It is also possible to consider generalizations corresponding to the interacting dimer model\cite{Ikhlef}, which have a radius $R\neq 1$. 
In terms of dimers a $\uparrow$ spin (or a fermion) is either an even vertical link occupied by a dimer, or an empty odd vertical link. The ferroelectric string in the spin language corresponds here to an alternating sequence $|1010\ldots 10\rangle$ of links occupied ($1$) by a dimer, and empty links ($0$). The EFP of both periodic chains can easily be calculated as a determinant $\mathcal{P}=\det_{0\leq i,j\leq \ell-1} (m_{ij})$ using Wick's theorem \cite{EFP_XX}. The matrix elements are given by
\begin{equation}
 m_{ij}=\frac{\sin \frac{\pi}{2}(i-j)}{L\sin \frac{\pi }{L}(i-j)}
\end{equation}
for the periodic XX chain at half-filling. In the limit $L\to \infty$ the corresponding symbol takes the form
\begin{equation}
 g(\phi)=\frac{1}{2}+\frac{1}{2}{\rm sgn}\left( \cos \phi\right)=\left\{
 \begin{array}{ccc}
  1&,& \pi\leq \phi \leq 3\pi/2\\
  0&,&{\rm otherwise}
 \end{array}
  \right.
\end{equation}
so that it vanishes on two whole intervals of $[0;2\pi]$. This is a big difference with the Ising case, as we cannot apply Fisher-Hartwig related results to this determinant. Here the enhanced scaling proportional to $\ell^2$ follows from Ref.~\cite{Widom_circular}. 
In the dimer chain the matrix elements are given by
\begin{eqnarray}
m_{ij}=\frac{\delta_{ij}}{2}+\frac{(-1)^{ij}}{L}\sum_{q\in Q}\frac{\cos \left[\frac{q}{2}(1+(-1)^{i-j})\right]}{\sqrt{1+\cos^2 q}}\cos[q(i-j)]
\end{eqnarray}
Here $Q=\Set{-\pi +(2m+1)\pi/L|m=0,\ldots,L/2}$. This matrix is only block Toeplitz, because of the factor $(-1)^{ij}$. There are also similar formulae for the open chains. Finally, note that such determinants are closely related to the so-called ``gap-probability'' encountered in the study of random matrices \cite{Dyson}. 
  \subsection{Imaginary time picture, and the arctic circle phenomenon}
  \label{sec:pictures}
Before presenting our results, let us mention that it is possible to look at the imaginary time picture for a weakly magnetized string, where bosonization applies \cite{AbanovKorepin}. It was shown that already in this limit, the slit shown in Sec.~\ref{sec:cbc} becomes an \emph{ellipse}, where the degrees of freedom are frozen. It was however noted that this picture agrees only qualitatively with the numerical data at maximum magnetization. Here we determine the region numerically for finite but large $\ell$ in the dimer model. The XX chain should give similar results.
\begin{figure}[htbp]
 \begin{subfigure}[b]{0.5\textwidth}
 \centering
\includegraphics[height=4.1cm]{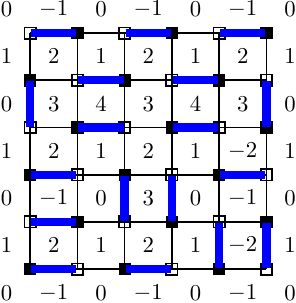}
  \caption{$6\times 6$ square lattice.}
  \label{fig:heights1}
 \end{subfigure}
 \begin{subfigure}[b]{0.5\textwidth}
 \centering
\includegraphics[height=4.1cm]{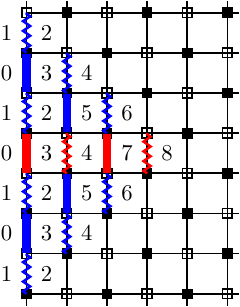}
\hfill
  \caption{Emptiness formation probability}
  \label{fig:heights2}
 \end{subfigure}
 \caption{Height mapping. (a): $6\times 6$ square lattice with open boundaries. (b): a sequence $|\!\uparrow\uparrow\uparrow\uparrow\rangle=|1010\rangle$ is imposed at the left of the fourth row in a semi-infinite grid. The dimers are the red thick lines, the empty links are the red zigzag lines. Due to the hardcore constraint, the occupation probabilities of several other links are set to $p=0$ or $p=1$ (in blue). The frozen region corresponds to a maximal slope for the height.
 }
 \label{fig:heights}
\end{figure}

We first recall the height mapping for dimers. Let us associate an integer height with each plaquette of the lattice, as illustrated in Fig.~\ref{fig:heights}.
The heights are defined as follows. Set the value on the lower leftmost face to be zero. Then, turning counterclockwise around a site of the even (resp. odd) sublattice represented by black squares (resp. white squares), the height picks $+3$ (resp. $-3$) when crossing a dimer, $-1$ (resp. $+1$) otherwise. It is easy to see that the flat configurations should dominate the others: after coarse-graining the long distance physics is well described by the gaussian action of Eq.~(\ref{eq:action}). 

However, the $EFP$ is exactly the probability of observing a ferromagnetic string such as the one shown in Fig.~\ref{fig:heights2} (in red). This corresponds to imposing the maximum slope on the height field (injecting lots of fermions in fermionic language). Such a constraint has far reaching consequences, due to the dimer hardcore rules. In particular, Fig.~\ref{fig:heights2} shows (in blue) all the links whose occupation probabilities are automatically set to $0$ or $1$. In this triangular region the dimers are completely frozen.
This observation already explains the scaling of the $\log EFP$: since there is a deficit in bulk free energy proportional to the area of this region, the $\log EFP$ will be proportional to $\ell^2$. For the six-vertex model the same argument applies, using the ice rules instead of the dimer hardcore constraint. Note also that the argument is so far independent of possible interactions, that can be introduced by a non-zero $\Delta$ in the XXZ chain, or dimer interactions on the plaquettes.

\begin{figure}[htbp]
\begin{center}
\begin{subfigure}[b]{0.46\textwidth}
\centering
 \includegraphics[width=6.05cm]{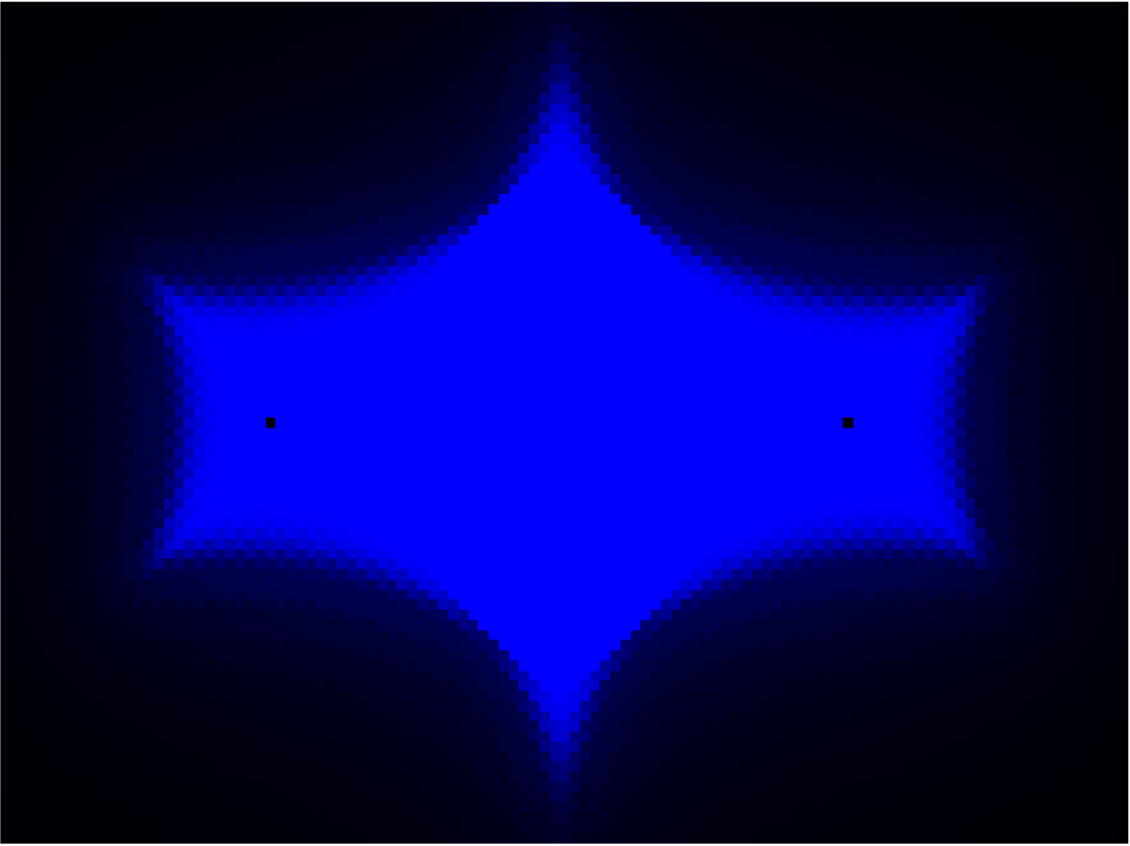}
 \caption{frozen region}
 \label{fig:arctic1a}
\end{subfigure}
\hfill
\begin{subfigure}[b]{0.23\textwidth}
\centering
 \includegraphics[width=3cm]{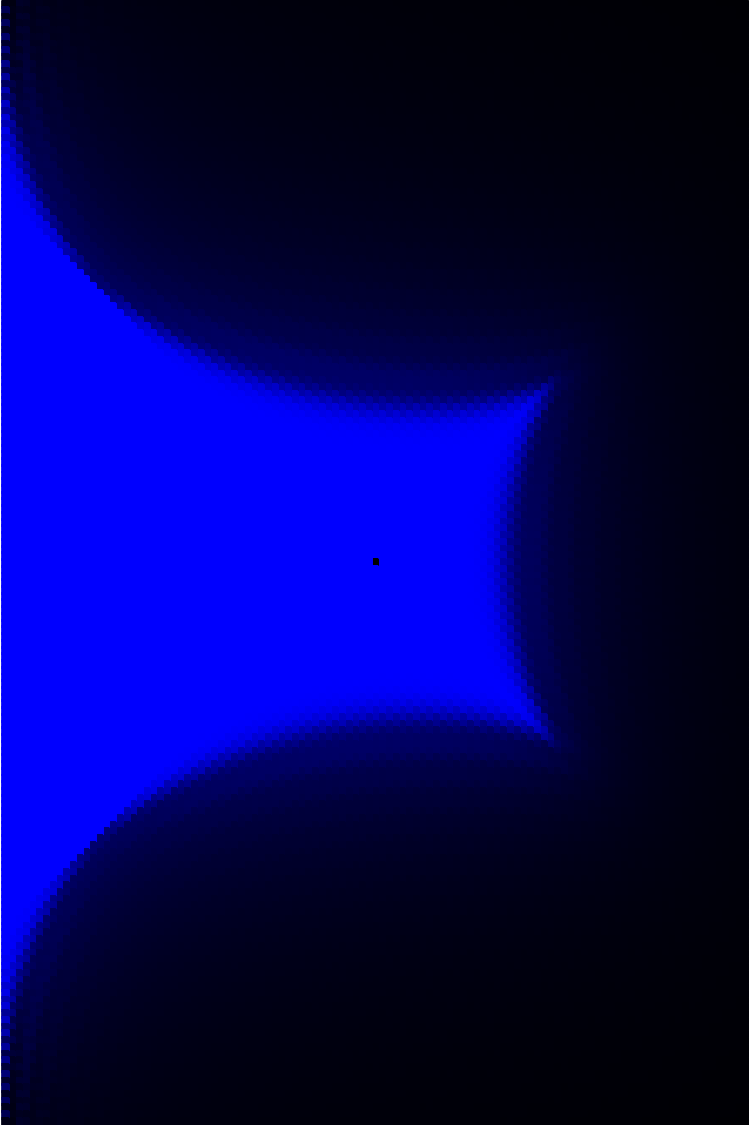} 
 \caption{frozen region}
 \label{fig:arctic1b}
\end{subfigure}
\hfill
\begin{subfigure}[b]{0.23\textwidth}
\centering
  \includegraphics[width=3cm]{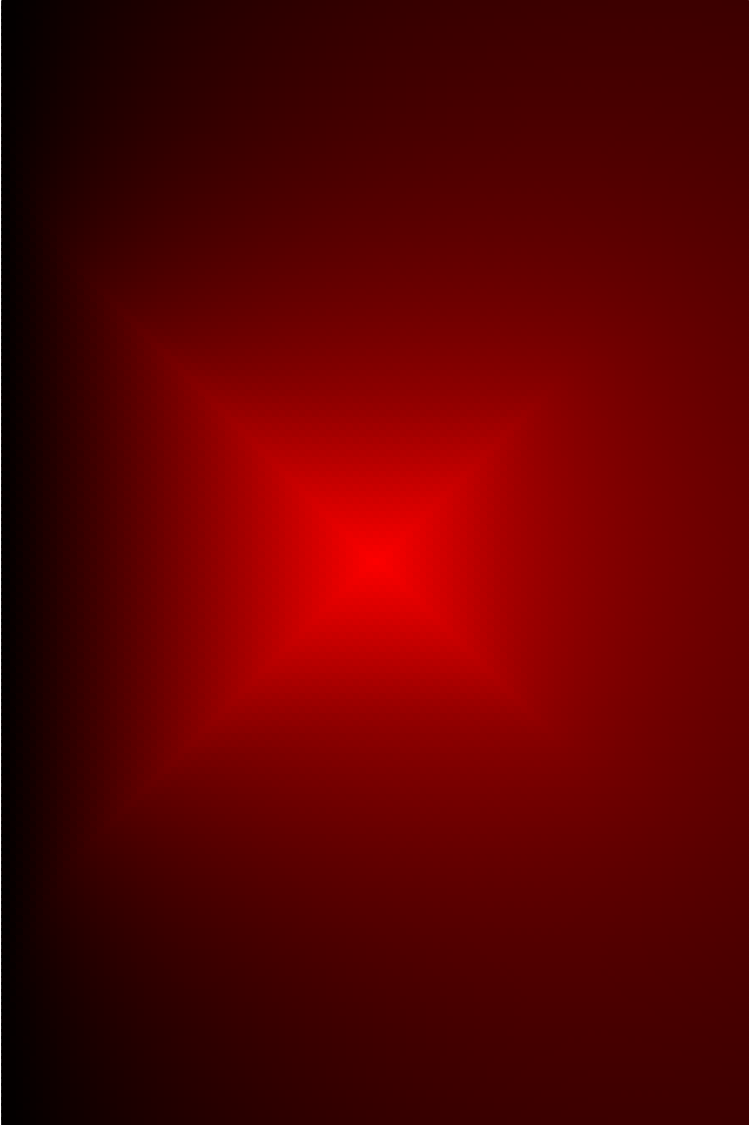}
  \caption{height profile}
  \label{fig:arctic1c}
\end{subfigure}
\end{center} 
\caption{Arctic phenomenon generated by the ferroelectric configuration $\uparrow\uparrow\ldots\uparrow$. (a): Slit of length $\ell=56$ in the infinite geometry. The data is shown on a $112\times84$ rectangle, and the frozen region is in blue. (b): Slit of length $\ell=56$ in the semi-infinite geometry. The data is shown on a $112\times 168$ rectangle. (c): Slit of length $\ell=56$ in the semi-infinite geometry. Same setup as for (b), but with the discrete heights shown (high heights are in red).}
\label{fig:arctic1}
\end{figure}

In the limit $\ell \to \infty$ such constraints become highly non trivial: the precise form of frozen region extends further than shown in Fig.~\ref{fig:heights2}, and should a priori depend on the interactions. It is represented in Fig.~\ref{fig:arctic1b}, for a slit of length $\ell=56$. We also show in Fig.~\ref{fig:arctic1a} the region corresponding to the infinite geometry.
The pictures are obtained as follows. First we calculate the probabilities associated to dimer occupancies on all links. The procedure to get them is described in \ref{app:dimers}, and makes extensive use of the Pfaffian solution \cite{Kasteleyn} of the dimer model. Second we associate a mean height to each plaquette using these probabilities\footnote{For example the height picks  $\pm\left(3\times p-1\times (1-p)\right)=\pm (4p-1)$ when crossing a link occupied by a dimer with probability $p$, and empty with probability $1-p$.}. Third we define a discrete squared gradient 
\begin{equation}
\left(\nabla_{\rm \!d} \varphi\right)^2=\left[\varphi(i_x,i_y+1)-\varphi(i_x,i_y-1)\right]^2+\left[\varphi(i_x+1,i_y)-\varphi(i_x-1,i_y)\right]^2.
\end{equation}
The maximum gradient is $ \left(\nabla_{\rm \!d} \varphi\right)^2=16$. We draw in blue the frozen region where the gradient is maximal and in black the regions with zero gradient. Intermediate regions are shown as a mixture of the two colors. Fig.~\ref{fig:arctic1a} (resp.~\ref{fig:arctic1b}) shows the results for the frozen region in the infinite (resp. semi-infinite) geometry. We also show in Fig.~\ref{fig:arctic1c} an example of height profile.

The limiting shape is a non trivial curve with several cusps. Notice that outside of the frozen region the dimer occupancies are still constrained (and therefore anisotropic), although they are not sufficiently far away. Using this we can interpret the leading term in the EFP: it is simply due to a deficit in free energy coming from the arctic region, as well its proximity, where the free energy is not constant with the position. Interestingly, the limiting shape for the semi-infinite geometry with slit length $\ell$ seems to match perfectly half that of the infinite geometry with slit $2\ell$. This observation is consistent with the exact results of \cite{EFP_Kozlowski,EFP_XXZrazumov}, which give $a_2({\rm semi-inf})=2a_2({\rm inf})$ in the XX chain. 

\begin{figure}[htbp]
 \begin{subfigure}[b]{0.32\textwidth}
 \centering
   \includegraphics[width=3cm]{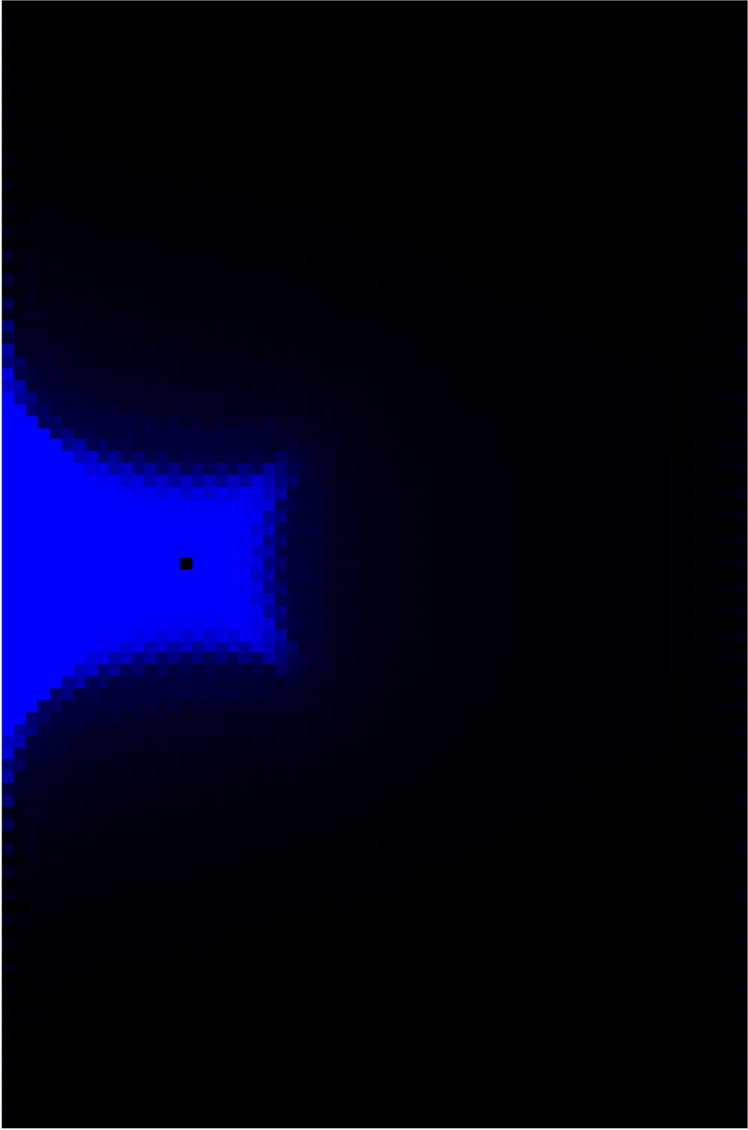}
  \caption{$\ell=16$}
  \label{fig:arctic_finitea}
 \end{subfigure}
 \begin{subfigure}[b]{0.32\textwidth}
 \centering
   \includegraphics[width=3cm]{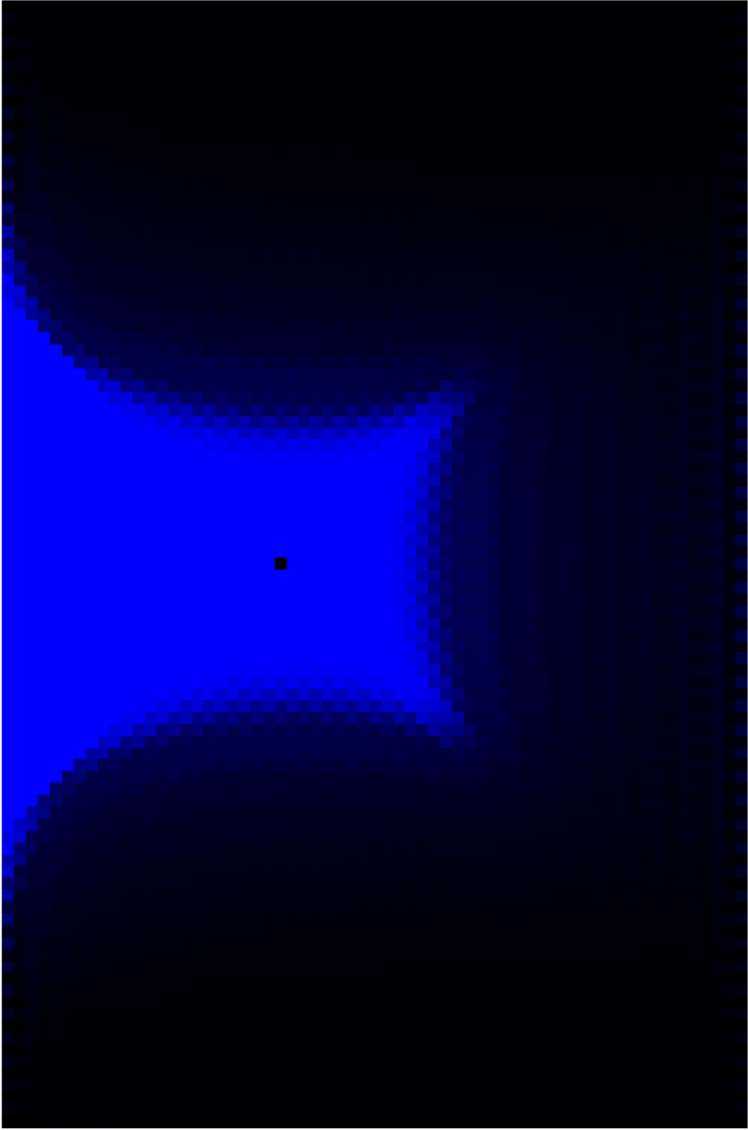}
  \caption{$\ell=24$}
  \label{fig:arctic_finiteb}
 \end{subfigure}
 \begin{subfigure}[b]{0.32\textwidth}
 \centering
   \includegraphics[width=3cm]{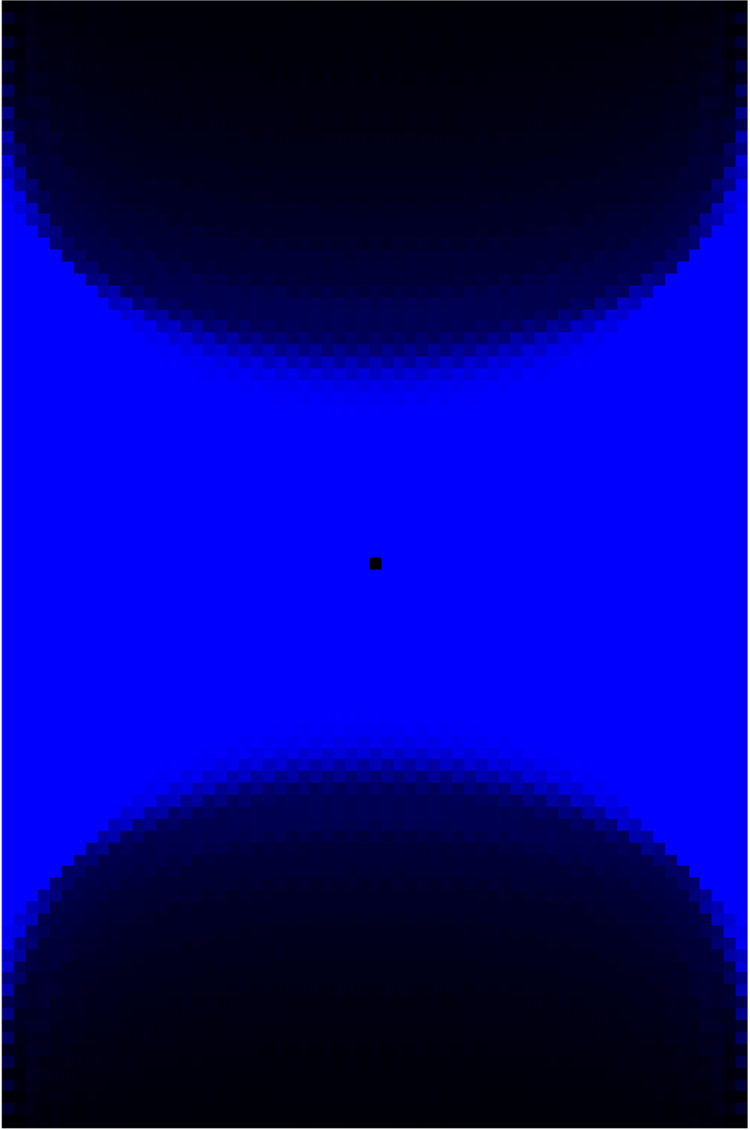}
  \caption{$\ell=32$}
  \label{fig:arctic_finitec}
 \end{subfigure}
 \caption{Arctic phenomenon for an open system of total length $L=64$. The frozen region is shown for three different slit lengths $\ell=16$ (a), $\ell=24$ (b) and $\ell=32$ (c). Notice for case (c) that the blue region cuts the system in two.}
 \label{fig:arctic_finite}
\end{figure}

Inspired by the results of Sec.~\ref{sec:cbc}, it is tempting to look at the arctic regions for a finite aspect ratio $\ell/L$, to see if the frozen region is modified. We present in Fig.~\ref{fig:arctic_finite} some numerical results regarding this question in the open geometry (the periodic geometry is similar). We observe that, as long as $\ell/L$ remains relatively small (figs.~\ref{fig:arctic_finitea} and \ref{fig:arctic_finiteb}), the picture looks more or less the same, up to a global dilatation. However the leading coefficient of the $\log EFP$ seems to change slightly (anticipating on table.~\ref{tab:numerics}), when varying the aspect ratio, so that this observation is probably only approximate. Also the limiting case $\ell/L=1/2$ is interesting, because the sequence of $L/2$ consecutive up spins has to be followed by $L/2$ consecutive down spins, due to the zero magnetization constraint. In this case the frozen region even cuts the imaginary time picture in
two disconnected parts (see figure.~\ref{fig:arctic_finitec}). It will be shown in Sec~\ref{sec:logterms} to have a huge impact on the logarithmic terms. 

\begin{figure}[htbp]
   \begin{subfigure}[b]{0.49\textwidth}
  \centering
    \includegraphics[width=5.32cm]{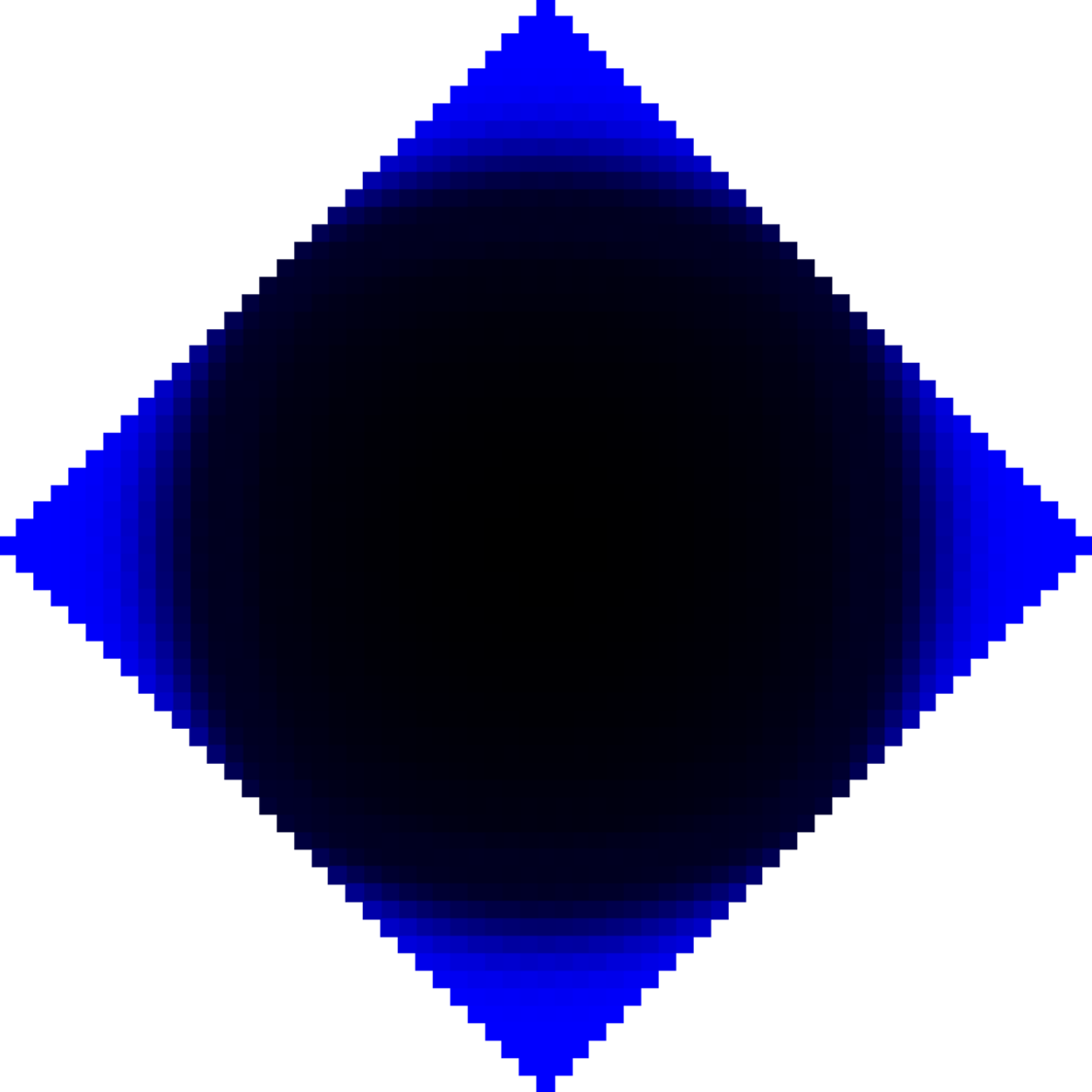}
\caption{Arctic circle on the Aztec diamond}
\label{fig:arctic_circlea}
\end{subfigure}
\begin{subfigure}[b]{0.49\textwidth}
 \centering
\includegraphics{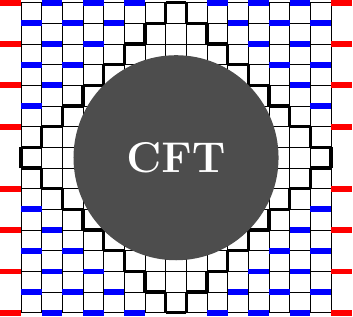}
\caption{Aztec diamond and emptiness formation probability}
\label{fig:arctic_circleb}
\end{subfigure}
  \caption{Arctic circle phenomenon. (a): Frozen region on a $32\times 32$ Aztec diamond. All dimer occupancies are set with probability one outside a circle in the thermodynamic limit. (b): The Aztec diamond (thick black lines) can also be obtained by imposing the boundary conditions $\uparrow\uparrow\uparrow\uparrow\uparrow\uparrow\uparrow\uparrow\downarrow\downarrow\downarrow\downarrow\downarrow\downarrow\downarrow\downarrow$ at the left and right of a $L\times L$ square (in red). In terms of dimer occupancies the boundary condition reads $1010101001010101$. Inside the circle lives a field-theory, whose exact properties are so far unknown.}
  \label{fig:arctic_circle}
\end{figure}

Such a behavior is yet another manifestation of the arctic phenomenon, first discovered in the study of dimers on the Aztec diamond \cite{JPS}. Compared to the standard rectangular geometry, the dimers are highly constrained near the corners of the diamond, and these constraints propagate on macroscopic scales. In terms of heights the slope is maximal on all sides of the diamond, and the authors of \cite{JPS} managed to prove that all degrees of freedom outside of the arctic circle were frozen. An illustration of this phenomenon is shown in figure.~\ref{fig:arctic_circlea}. 
Note that the Aztec diamond can easily be generated by imposing maximally magnetized boundary conditions on a square: in terms of transfer matrix the partition function for dimers on the Aztec diamond of diameter $L_x$ reads
\begin{equation}
 Z=\langle\uparrow\ldots \uparrow\downarrow\ldots\downarrow |T^{L_x}|\uparrow\ldots \uparrow\downarrow\ldots\downarrow\rangle,
\end{equation}
provided $L_y\geq L_x$. $T$ is the transfer matrix of the dimer model, acting on the vector space of dimension $2^{L_y}$ generated by the dimer occupancies along a vertical line.
  See figure.~\ref{fig:arctic_circleb} for an explicit example  in the case $L_x=L_y=16$. 

Let us finally mention that the arctic curve separating the frozen region from the fluctuating (massless) region can usually be understood in  terms of a entropy minimization principle \cite{Kenyon1,Kenyon2}. It would be interesting to apply this method to our problem, and attempt an exact calculation of the regions shown in figures \ref{fig:arctic1a} and \ref{fig:arctic1b}. 
\subsection{Logarithmic terms}
\label{sec:logterms}
As for the conformal boundary case, the logarithmic terms in the EFP are potentially the most interesting, as it is reasonable to assume they might be universal. Many exact results are already available, they are summarized in table.~\ref{tab:coefslogs} for the XXZ chain. We also show the terms for the six-vertex analog of dimers on the Aztec diamond, namely the six-vertex model with domain wall boundary condition (DWBC) \cite{dwbc1,dwbc2,dwbc3}.
\renewcommand{\arraystretch}{1.5}
\begin{table}[htbp]
\centering
\begin{tabular}{|c|c|c|c|}
   \hline $\Delta$ & $b_0(\Delta)$, periodic geometry & $b_0(\Delta)$, open geometry & six-vertex DWBC\\
   \hline $0$ & $1/4$ &$1/8$&0\\
    $1/2$ & $5/36$ & $-1/72$&$-1/18$\\
    \hline
    general & $\frac{1}{12}\left(2R^2+8/R^2-7\right)$ &  ?&$\frac{1}{12}\left(R^2+4/R^2-5\right)$\\
    \hline
  \end{tabular}
  \caption{Summary of the logarithmic terms in the scaling (\ref{eq:xxzscaling}) of the $\log EFP$ for the XXZ chain. The general result for $b_0(\Delta)$ in the periodic geometry has been conjectured and numerically checked in \cite{KorepinLukyanov}. The others are exact \cite{EFP_XX,EFP_XXZrazumov,EFP_Kozlowski,dwbc_as1,dwbc_as2}.}
  \label{tab:coefslogs}
\end{table}
The prefactor of the logarithm is always given by a simple function of the radius $R$, and does not depend on non universal features such as the Fermi-speed, etc. This is an additional hint for universality, but needs to be checked more carefully. In particular, we have seen in Sec.~\ref{sec:pictures} that bozonization breaks down, and it is not a priori obvious whether $R$ can still be interpreted as the compactification radius of the Luttinger liquid.

The non frozen degrees of freedom have non trivial long-range correlations, and should still be described by a massless field theory. However it is clearly neither translational invariant nor scale invariant, and this makes it much more difficult to apprehend than a CFT. The simplest scenario would be that the action (\ref{eq:action}) still describe the low energy fluctuating degrees of freedom, albeit with a position-dependent stiffness $\kappa(x,y)$. Alternatively, this can be seen as a field-theory in a curved space-time, whose flat space-time analog is conformally invariant. The advantage of such a picture is that the radius $R$ still makes sense, as the compactification radius of the flat space-time analog theory. A possible way to check this would be to exploit some exact results for the dimers Green's function on the Aztec diamond \cite{Helfgott}, and determine this way the asymptotic behavior of correlations anywhere inside the diamond. However this might not be so straightforward, and falls outside the scope of the present paper.

Such logarithmic terms can also be compared with those appearing on a rectangle, which are simpler from the field-theoretical point of view. There is however a subtlety that needs to be taken into account. Indeed using the height mapping, it is easy to see that the boundary heights are not exactly the same on the two sides touching a given corner. See figure.~\ref{fig:heights1}, that shows a coarse grained height $\langle h\rangle=\pm 1/2$ depending on the side. This small mismatch has a tremendous impact in the thermodynamic limit, and needs to be dealt with by adding a harmonic function in the action, that encodes the height shift \cite{Stephan_phasetransition}. In the language of boundary CFT, this corresponds to the insertion of $4$ (magnetic vertex) boundary changing operators  \cite{Cardybcc}. It is then straightforward to apply the general analysis of Ref.~\cite{Rectangle2}, and one even gets this way the full partition function
\begin{equation}\label{eq:rect_cft}
 \mathcal{Z}_{\rm rect}={\rm cst}\times L_x^{c/4-R^2/4}\frac{\left[\theta_\nu(\tau)\right]^{R^2/2}}{\left[\eta(\tau)\right]^{c/2}}\qquad,\qquad \tau=i\frac{L_y}{L_x},
\end{equation}
with central charge $c=1$. Note the non-universal constant prefactor. Here $\nu$ depends on the parity of $L_y$: $\nu=3$ for even $L_y$ and $\nu=4$ for odd $L_y$\footnote{The difference when $L_y$ is odd is related to the order in which the height shifts are applied: we have the boundary correlator $\langle V_{\alpha}(z_1)V_{-\alpha}(z_2)V_{\alpha}(z_3)V_{-\alpha}(z_4)\rangle$ (resp. $\langle V_{\alpha}(z_1)V_{\alpha}(z_2)V_{-\alpha}(z_3)V_{-\alpha}(z_4)\rangle$) for $L_y$ even (resp. odd). Here $z_1$ (resp. $z_2$, $z_3$, $z_4$) denotes the position of the bottom left (resp. top left, top right, bottom right) corner, and $\alpha^2=R^2/16$ \cite{Stephan_phasetransition}. See \cite{Stephan_rvb} for a similar even-odd effect on the cylinder.}. At the free fermion point $R=1$, and Eq.~(\ref{eq:rect_cft}) reproduces the exact result of Ferdinand \cite{Ferdinand}. $\theta_3$ and $\theta_4$ are the third and fourth Jacobi theta functions, given by
\begin{eqnarray}
 \theta_3(\tau)&=&\sum_{k \in \mathbb{Z}}e^{i\pi k^2\tau},\\
 \theta_4(\tau)&=&\sum_{k \in \mathbb{Z}}(-1)^k e^{i\pi k^2\tau},
\end{eqnarray}
and $\eta$ is the Dedekind eta function
\begin{equation}
 \eta(\tau)=e^{-i\pi \tau/12}\prod_{k=1}^{\infty}\left(1-e^{2i\pi k \tau}\right).
\end{equation}
Notice the similarity of the logarithmic term $(R^2/4-1/4)\log L$ in Eq.~(\ref{eq:rect_cft}), with those summarized in table.~\ref{tab:coefslogs}. In addition to the logarithmic terms, Eq.~(\ref{eq:rect_cft}) also shows that the shape dependence of the free energy is universal for the dimer model, and this also applies to the six-vertex model. 

\textcolor{black}{Let us now come back to the emptiness formation probability. We study it numerically in the periodic XX and dimer chains for various aspect ratios $x/\ell$, and try to identify which terms are universal.}
The results for a periodic chain at half-filling ($\rho=1/2$) are shown in table.~\ref{tab:numerics}.
\begin{table}[htbp]
 \centering
 \begin{tabular}{|c|c|c|c|c||c|c|c|c|}
 \hline &\multicolumn{4}{c||}{XX chain ($\rho=1/2$)}&\multicolumn{4}{c|}{dimer chain ($\rho=1/2$)}\\
  \hline coef & $x=\rho/4$& $x=\rho/2$& $x=2\rho/3$&$x=\rho$ & $x=\rho/4$& $x=\rho/2$&$x=2\rho/3$ &$x=\rho$ \\
  \hline $a_{2}$ &0.3498456&0.3604522&0.3729935&0.4262783&0.3984028&0.4139014&0.4321317&0.5047001\\
  \hline $b_0$ &0.2500058&0.2500012&0.2499928&0.1666604&0.2500008&0.2499955&0.2499783&0.1666236\\
  \hline $a_{0}$ &0.3468585&0.3297074&0.3065721&0.2556067&0.3295512&0.3230345&0.3161439&0.3289082\\
\hline
 \end{tabular}
 \caption{Scaling of the $\log EFP$ in the periodic XX and dimer chains, for various aspect ratios $x=\ell/L$, at half-filling $\rho=1/2$. The data is extracted via a fit to the form $a_2\ell^2+b_0\log \ell+a_0$ for system sizes between $\ell=80$ and $\ell=96$.}
 \label{tab:numerics}
\end{table}
We observe that the leading contribution $a_2$ depend on the ratio $x=\ell/L$ and differ for the two chains. However, the logarithmic terms $b_0\simeq 1/4$ stand compatible with table \ref{tab:coefslogs}, as long as the aspect ratio is smaller than $x=1/2$. The precise point $x=1/2$ corresponds to the case shown in Fig.~\ref{fig:arctic_finitec}, where the $L/2$ up spins have to be followed by $L/2$ down spins; the coefficient has a jump to a value very likely to be $1/6$. For $x>1/2$ the EFP is zero. Let us also mention that the precision on $b_0$ can be increased 
by adding subleading corrections in the form of a power series in $1/L$. For example going up to $1/L^2$ we find $|b_0-1/4|<10^{-8}$ for all the corresponding values in table.~\ref{tab:numerics}. Finally, we looked at the subleading constant, that is not expected to be universal. However, differences between subleading constants for different aspect ratios have been shown to be universal for Ising (see Sec~\ref{sec:logarithmic_terms}), and also for vertex/dimers on the rectangle (see Eq.~\ref{eq:rect_cft}). This is not the case here: as can be seen the values taken by $a_0(x)-a_0(x^\prime)$ for different $x$ and $x^\prime$ differ between the $XX$ and dimer chains. Therefore, $b_0$ is the only term that can be universal.  

\textcolor{black}{We also performed numerical computations away from half-filling. This can be obtained by looking at a sector with a different fermion number $N=\rho L$ with $\rho \neq 1/2$, or as the true ground-state in a magnetic field. 
 The data for $\rho=1/4$ is shown in table.~\ref{tab:hnumerics}.}
 \begin{table}[htbp]
 \centering
 \begin{tabular}{|c|c|c|c|c||c|c|c|c|}
   \hline &\multicolumn{4}{c||}{XX chain ($\rho=1/4$)}&\multicolumn{4}{c|}{dimer chain ($\rho=1/4$)}\\
 \hline coef & $x=\rho/4$& $x=\rho/2$& $x=2\rho/3$&$x=\rho$ & $x=\rho/4$& $x=\rho/2$&$x=2\rho/3$ &$x=\rho$ \\
  \hline $a_{2}$ &0.9652946&0.9804121&0.9977718&1.0656959 & 1.1015673&1.1189505&1.1386701&1.2133840\\
  \hline $b_{0}$ &0.2500061&0.2500080&0.2500150&0.1666715 & 0.2500055&0.2500076&0.2500146&0.1666716\\
  \hline $a_{0}$ &0.4134223&0.3952091&0.3705907&0.3133143 & 0.4069374&0.3952914&0.3775115&0.3397236\\
  \hline
 \end{tabular}
 \caption{Scaling of the $\log EFP$ in the periodic XX and dimer chains, for various aspect ratios at quarter-filling $\rho=1/4$. Same extraction procedure as in tables.~\ref{tab:numerics}.}
 \label{tab:hnumerics}
\end{table}
\textcolor{black}{We observe that the logarithmic terms remain the same, $b_0=1/4$ for $x<\rho$. For $x=\rho$ a similar phenomenon as the one shown in Fig.~\ref{fig:arctic_finitec} occurs. In the $XX$ and dimer chains the compactification radius does not depend on the magnetic field, so this result is additional evidence for the universality of $b_0$. It is also possible to look at a deformation of the dimer chain corresponding to the addition of different fugacities for horizontal and vertical dimers. This changes the dispersion relation but not the universality class, and we found the same values as in tables \ref{tab:numerics} and \ref{tab:hnumerics} for $b_0$. 
}

Finally, we performed a similar study for all these chains with open boundary conditions. The numerical computations give $b_0=1/8$ for $x<\rho$ and $b_0=1/12$ at $x=\rho$, namely half those in the periodic chain. All these results provide strong evidence for the universality the $EFP$. It would be interesting to perform a similar study for other models with interaction, such as the XXZ chain with general $\Delta$, interacting dimers \cite{Ikhlef}, or a $J1-J2$ chain \cite{J1J2}. The latter two would be most interesting, as they are non-integrable and rigorous derivations of the asymptotic expansions would be out of reach.

\section{Some related problems}
 \label{sec:other}
 We discuss in this section some possible extensions of this work, as well as some related questions. 
\subsection{The other Ising critical line in the XY chain}
\label{sec:EFP_down}
This line with $\gamma=1$ and $h=-1$ also belongs to the Ising universality class, and has also been studied by Franchini and Abanov \cite{FranchiniAbanov}. They showed, in the infinite geometry, that the $\log EFP$ scales as
\begin{equation}\label{eq:EFP_scaling_down}
 \mathcal{E}_p=a_1\ell+\frac{1}{16}\log \ell +a_0^{\prime (p)}+\frac{a_{-1/2}}{\sqrt{\ell}}+ \ldots
\end{equation}
The slightly different scaling is mathematically explained by the fact that the symbol
\begin{equation}
 g(\phi)=\frac{1}{2}-\frac{1}{2}\exp\left(-i\arctan\left[\nu \tan \frac{\phi}{2}\right]\right)
\end{equation}
has an additional singularity at $\phi=0$, and subleading representations have to be taken into account (see Eq.~(\ref{eq:gfh})). Our analysis can in principle be extended to explain this scaling. First, notice that the EFP in this chain ($h=-1$) can be recast as the probability of observing a sequence 
$|\!\downarrow\downarrow\ldots\downarrow\rangle$ in the $h=1$ chain. Then, we have
\begin{eqnarray}
 |\!\downarrow\downarrow\ldots\downarrow\rangle_z&=&\frac{1}{2^{\ell/2}}\, \left(|\!\rightarrow\rangle_x-|\!\leftarrow\rangle_x\right)
  \left(|\!\rightarrow\rangle_x-|\!\leftarrow\rangle_x\right)
  \ldots  \left(|\!\rightarrow\rangle_x-|\!\leftarrow\rangle_x\right)\\
  &=&\frac{1}{2^{\ell/2}}\sum_{\{\sigma_j^x=\rightarrow,\leftarrow\}} (-1)^{N_{\leftarrow}} |\sigma_1^x \sigma_2^x\ldots \sigma_\ell^x\rangle\\
  &\neq &|{\rm free}\rangle_x,
\end{eqnarray}
where $N_\downarrow$ counts the number of down spins. Since the free boundary condition is unstable under the RG flow, this should actually renormalize to a (stable) \emph{fixed} boundary condition, and we have to introduce \emph{boundary changing operators}. In the infinite geometry they do not affect the Cardy-Peschel term because the external boundaries are at infinity, but they influence a lot the subleading corrections. For example we still expect a $\ell^{-1}\log \ell$ term to appear in (\ref{eq:EFP_scaling_down}), but with a prefactor modified to 
\begin{equation}
 b_{-1}=-\frac{\xi}{8\pi} \left(c-16h_{{\rm f}+}\right),
\end{equation}
where $h_{{\rm f}+}=1/16$ is the dimension of the operator changing the boundary condition from free to $+$. We refer to \cite{SD_2013} for the method. The changes are even more dramatic in the semi-infinite geometry. Applying the Cardy-Peschel formula (\ref{eq:CardyPeschel}) or taking the limit $L\to \infty$ in (\ref{eq:openbcc}), we get
\begin{equation}
 \mathcal{E}_o=a_1\ell+\left(4h_{{\rm f}+}-\frac{c}{16}\right)\log \ell +a_0^{\prime (o)}+o(1),
\end{equation}
so that the prefactor of the logarithmic term is modified to $7/32$. To our knowledge this exponent cannot be obtained using known theorems on Toeplitz+Hankel determinants. We also expect the presence of a $\ell^{-1}\log \ell$ correction among the subleading terms. 
\subsection{Full counting statistics}
\label{sec:fcs}
Another closely related problem is that of the counting statistics of magnetization in spin chains \cite{xzmagn}. What makes this type of problem interesting is that it is motivated by recent experiments in cold-atom systems. In particular, it is sometimes possible to use the quantum noise to measure the full probability distribution of fluctuating observables \cite{Gritsevetal,zmagn}. 

For exact studies it is usually more convenient to study the magnetization in the basis of the eigenstates of the $\sigma^z$ (see, however, Refs.~\cite{SokalSalas,xmagn}). The statistics can be conveniently expressed in terms of the following generating function
\begin{equation}\label{eq:generating}
 \chi_\ell(\lambda)=\Braket{ e^{-\lambda \sum_{j=1}^{\ell}(1-\sigma_j^z)/2}}=\sum_{m\geq 0} p_m e^{-m\lambda}, 
\end{equation}
where $p_m$ is the probability of having $m$ spins down (out of $\ell$) in the z-basis. As we shall see, $\chi_\ell(\lambda)$ is a natural generalization of the EFP. All cumulants are recovered as
\begin{equation}
 \Braket{m^p}_c=(-1)^p\left.\frac{d^p \log \chi_\ell(\lambda)}{d\lambda^p}\right|_{\lambda=0}
\end{equation}
Here we look at the XY chain in transverse field (\ref{eq:xychain}) on the Ising critical line. 
Using the free fermion structure \cite{LiebSchultzMattis}, the generating function can be recast as a Pfaffian, which simplifies into a Toeplitz determinant $\chi_\ell(\lambda)=\det ([g_\lambda]_{i-j})$. The symbol is (see e.g. \cite{zmagn,AbanovIvanov})
\begin{equation}
 g_\lambda(\phi)=\frac{1+e^{-\lambda}}{2}+\frac{1-e^{-\lambda}}{2}\exp\left(-i\arctan \left[\nu \tan \frac{\phi}{2}\right]\right).
\end{equation}
Next we apply the analysis of Sec.~\ref{sec:toeplitz_subleading}, limiting ourselves to $e^{\lambda}\in [-1;+\infty]$ for simplicity. The symbol can be put in the form (\ref{eq:new_parametrization}), with $\beta=-\arctan[\tanh (\lambda/2)]/\pi$ and $\eta_1=-\tanh \lambda/(4\pi \nu)$. The asymptotic expansion reads
\begin{eqnarray}
 -\log \chi_\ell(\lambda)=a_1(\lambda)\ell+b_0(\lambda) \log \ell+a_0(\lambda)+b_{-1}(\lambda)\ell^{-1}\log \ell+\mathcal{O}(\ell^{-1}),
\end{eqnarray}
and the coefficients are given by
\begin{eqnarray}
a_1(\lambda)&=&[\log \left|g_\lambda\right|]_0\\
a_0(\lambda)&=&-\sum_{k=1}^{\infty} k \left([\log \left|g_\lambda\right|]_k\right)^2-\log \left(G(1+\beta)G(1-\beta)\right)\\
b_0(\lambda)&=&\frac{1}{\pi^2}\arctan^2\left[\tanh \frac{\lambda}{2}\right]\\\label{eq:bm1}
 b_{-1}(\lambda)&=&-\frac{b_0(\lambda)}{2\pi \nu}\tanh \lambda
\end{eqnarray}
$a_1(\lambda)$ was studied over the unit circle $|e^{\lambda}|=1$ in \cite{AbanovIvanov}, and $a_0(\lambda),b_0(\lambda)$ are extensions for $\nu\neq 1$ of the results of Ref.~\cite{zmagn} for the ICTF ($\nu=1$). The last term $b_{-1}(\lambda)$ is new. 

Two special values of the generating function are of interest, in addition to the obvious $\chi_\ell(0)=1$. First, $\lim_{\lambda\to \infty}\chi_\ell(\lambda)$ is nothing but the emptiness formation probability, and we recover $b_0=1/16=c/8$, as well as $b_{-1}=-1/(32\pi \nu)=-\xi c/(8\pi)$. Second, one can show, using a Kramers-Wannier transformation, that \cite{zmagn}
\begin{equation}
 \chi_\ell(2i\pi)\propto \Braket{\sigma_1^x\sigma_\ell^x}\sim \ell^{-1/4}
\end{equation}
and one recovers the result of \cite{Pfeuty,Barouch}. In CFT language the exponent is twice the scaling dimension ($1/8$) of the Ising spin field. Note that it is possible to interpret the $\lambda$ dependence of $b_0(\lambda)$ and $b_{-1}(\lambda)$ within CFT. In terms of the Majorana fermion fields, the generating function can be expressed as
\begin{equation}
\chi_\ell(\lambda) =\Braket{\exp\left(-i\lambda \int_0^{\ell}dx\, \psi(x)\bar{\psi}(x)\right)}.
\end{equation}
where $\psi$ and $\bar{\psi}$ are the analytic and anti-analytic components of the Majorana fermion. 
The argument of the exponential can be seen as a perturbation of the CFT action, and has been studied in Ref.~\cite{FendleyFisherNayak}. The imaginary time pictures shown in Fig.~\ref{fig:euclidean} remain valid. There is however an important subtlety: the term $\int_0^{\ell}dx\, \psi(x)\bar{\psi}(x)$ is marginal along the slit, so that the conformal boundary condition are not free anymore. To determine them properly, one can fold the model about the slit \cite{OshikawaAffleck}, and then bosonize the two resulting Ising copies. The corresponding boundary condition, for a given $\lambda>0$, is then part of the one-parameter family of continuous Dirichlet boundary condition investigated in Ref.~\cite{OshikawaAffleck}. We refer to to \cite{FendleyFisherNayak,OshikawaAffleck} for a thorough discussion. 

From the generating function it is possible to reconstruct the cumulants. For example the first three in the ICTF are given by
\begin{eqnarray}
  \langle m \rangle_c&=&\left(\frac{1}{2}-\frac{1}{\pi}\right)\ell \\
  \langle m^2 \rangle_c&=&\frac{1}{4}\ell-\frac{1}{2\pi^2}\log \ell+{\rm cst}+\mathcal{O}(\ell^{-1})\\
  \langle m^3 \rangle_c&=&\frac{2}{3\pi}\ell+{\rm cst}-\frac{3}{4\pi^3}\frac{\log \ell}{\ell}+\mathcal{O}(\ell^{-1})
\end{eqnarray}
The linear scaling of all the cumulants means that the probability distribution $p_m$ is gaussian. Note that $b_0$ (resp. $b_{-1}$) is symmetric (resp. antisymmetric) under to $x\to x^{-1}$, hence $\log \ell$ (resp. $\ell^{-1}\log \ell$) terms can only appear in even (resp. odd) order cumulants.  

It is useful to compare this to the XX chain case (see e.g \cite{KlichYale2,Abanov_stats1}). Here an important difference is that the magnetization is conserved, and the system is equivalent to free fermions without pairing. The symbol reads
\begin{equation}
 g_\lambda(\phi)=\frac{1+e^{-\lambda}}{2}+\frac{1-e^{-\lambda}}{2} {\rm sgn}\left( \cos \phi-\cos k_F\right)
\end{equation}
and has two Fisher-Hartwig singularities at $\phi=k_F$ and $\phi=2\pi-k_F$. $k_F$ is the Fermi momentum. The corresponding generating function is given by
\begin{equation}
 -\log \chi_\ell (\lambda)=\frac{\lambda k_F}{\pi}\ell+\frac{\lambda^2}{2\pi^2}\log \ell+\mathcal{O}(1),
\end{equation}
with subleading corrections taking the form (\ref{eq:kozlowski_conjecture}). The result becomes singular when $\lambda \to \infty$; we have seen that in this limit $-\log \chi_\ell$ is proportional to $\ell^2$, not $\ell$. This is an important difference with the Ising case, as the counting statistics and the EFP have here a very different behavior. The cumulants are given by
\begin{eqnarray}
 \braket{m}_c&=&\frac{k_F}{\pi}\ell \\
 \braket{m^2}_c&=&\frac{1}{\pi^2}\log \ell+\mathcal{O}(1)
\end{eqnarray}
The higher-order cumulants $\braket{m^p}_c$ are either zero for $p$ odd or $\mathcal{O}(1)$ for $p$ even. All these can also be understood from simpler bosonization arguments. Interestingly, the even order cumulants can also be used to reproduce exactly the entanglement entropy \cite{KlichLevitov,KlichYale1,KlichYale2}, and therefore $S=(\pi^2/3)\braket{m^2}_c$ at the leading order.
Note that while this connection can be generalized to fermions with pairing (that describe the Ising chain or superconducting states), it is necessary to count the fermionic-quasiparticles appearing in the entanglement Hamiltonian \cite{KlichYale1,KlichYale2}, not the original fermions as we did in (\ref{eq:generating}). Hence the result we have established for the Ising chain does not seem to have any relation to the generating function whose even-order cumulants reproduce the entanglement entropy. This observation is consistent with the general results for the corrections to scaling \cite{ee_corrections_bulk,ee_corrections_boundary}.

 \subsection{Classical Mutual information in a spin chain}
 \label{sec:mutualinf}
 Let us consider the same bipartition as before for a spin-1/2 chain. Since ${\rm Tr}\, \rho_A=1$, each diagonal element $\braket{\sigma| \rho_A|\sigma}$ of the reduced density matrix  may be viewed as a probability. It is then natural to consider the classical R\'enyi-entropy of these probabilities
 \begin{equation}
  S_n(L,\ell)=\frac{1}{1-n}\ln \left(\sum_{\sigma}\left[\Braket{\sigma| \rho_A|\sigma}\right]^n \right),
 \end{equation}
 from which the classical Shannon entropy is a limiting case $n\to 1$. Here the sum runs over all $2^\ell$ possible spin configurations in subsystem $A$. This entropy is not to be confused with the R\'enyi entanglement entropy (REE), as the reduced density matrix is not diagonal. However when $\ell=L$, it is nothing but the REE of two-dimensional Rokhsar-Kivelson states in a cylinder/strip geometry \cite{Stephan09}, and has been further studied for this reason \cite{Stephan_Ising,Zaletel,Stephan_phasetransition}. For $\ell\neq L$ the correspondence doesn't apply anymore, but most of the techniques developed previously can still be applied. In particular, the general scenario of Refs.~\cite{Stephan_Ising,Stephan_phasetransition}, that predicts two different behaviors as a function of $n$ separated by a phase transition, should still hold. This is interesting because in the Ising chain the limit $n\to \infty$ of this entropy is nothing but the logarithmic emptiness formation probability. 
  Before presenting some possible consequences of this connection, it is convenient to introduce the R\'enyi Mutual information
\begin{equation}
 I_n=S_n(L,\ell)+S_n(L,L-\ell)-S_n(L,L).
\end{equation}
This combination filters out the line free energy contribution, so that the universal logarithms are now the leading terms. It was shown that for $n>1$, all universal properties are dominated by the $n\to\infty$ limit \cite{Stephan_Ising}. Using twice Eq.~(\ref{eq:main1}) we predict
\begin{equation}\label{eq:mutual}
 I_n=\frac{c}{4}\times \frac{n}{n-1}\log \left(\frac{L}{\pi}\sin \frac{\pi \ell}{L}\right)+{\rm cst.}+o(1)
\end{equation}
for a periodic chain, and a similar result in the open chain. The limit $n=1$ in the periodic case was studied numerically in Refs.~\cite{Hinrichsen,Grassberger}, and this is a more challenging problem in CFT. However, we notice that the conformal scaling of Eq.~(\ref{eq:mutual}) still seems to apply, albeit with a different and non trivial universal prefactor \cite{Hinrichsen,Grassberger} (see also Eq.~(III.181) in Ref.~\cite{StephanPhD} for an even more precise determination in the limit $L\to \infty$). Similar mysterious numbers have been observed numerically for the Shannon entropy of a full chain \cite{Stephan_Ising}. It would also be interesting to study in details the subleading corrections at $n>1$ and $n=1$, to see if the unusual $L^{-1}\log L$ term is still present as a subleading correction. 

For the XXZ and dimer chains the limit $n\to \infty$ doesn't give the emptiness formation probability, but rather the probability of observing an \emph{antiferromagnetic} string $p(\uparrow\downarrow\ldots\uparrow\downarrow)$. At the free fermions point the corresponding determinants are also simple and take the  form $p=\det_{1\leq i,j\leq \ell}\, (m_{ij})$ with  
 \begin{equation}
 m_{ij}=
 (-1)^{i\delta_{ij}}\frac{\sin\left[\frac{\pi}{2}(i-j)\right]}{L\sin\left[\frac{\pi}{L}(i-j)\right]}
 \end{equation}
for the XX chain and
\begin{equation}
 m_{ij}=\frac{\delta_{ij}}{2}+\frac{(-1)^{i(j+1)}}{L}\sum_{q\in Q}\frac{\cos \left[\frac{q}{2}(1+(-1)^{i-j})\right]}{\sqrt{1+\cos^2 q}}\cos[q(i-j)]
\end{equation}
for the dimer chain. Once again $Q$ is the set $Q=\Set{-\pi +(2m+1)\pi/L|m=0,\ldots,L/2}$. Here we have only given the determinants corresponding to a periodic chain, but there are similar expressions in the open case. It might be possible to study exactly the infinite limit $L\to \infty$, but the situation is more complicated because the corresponding matrices are only block-Toeplitz. Although some rigorous asymptotic results in such cases are available  \cite{blockT1,blockT2}, to the author's knowledge none of these can be applied to derive the universal terms as of now. It is however possible to perform a CFT analysis similar to the one presented in Sec.~\ref{sec:cbc}. Indeed imposing this antiferromagnetic sequence does not inject particles in the system, and the arguments developed in Sec~\ref{sec:cbc} should still apply. For example the leading logarithmic terms can easily be predicted in both chains. Since the configuration with alternating spins is expected to renormalize to a Dirichlet boundary condition for the boson field, we predict a term in $\case{c}{8}\log \ell$ in the infinite geometry. In the semi-infinite geometry one has to take care of the height shift discussed
in Sec.~\ref{sec:logterms}, and one arrives at $\case{2R^2-c}{16}\log\ell$. 
Here the central charge is $c=1$. However just like in Sec.~\ref{sec:EFP_down}, subleading corrections require a new and more thorough analysis, as other leading irrelevant operators could play a role. This is also an interesting question for further studies.  
 \section{Conclusion}
 We have studied in this paper the universal properties of the emptiness formation probability, mostly focusing on a CFT approach but also trying to combine it with other numerical and exact methods. It has been known for some time that the EFP can exhibit strikingly different behavior, depending on the model (e.g. Ising versus  XXZ).  
 We have shown how the Ising universality class can be systematically understood, and performed extensive checks of our results in a simple free fermionic realization. It however applies away from free fermions, and can also be generalized to any minimal model. 
 
 We used extensively the connection with the theory of Toeplitz and Toeplitz+Hankel determinants. While the application of such Fisher-Hartwig related results has become increasingly frequent in the physics literature, it usually allows to go further than field theoretical calculations. Here we managed to find an example where the CFT approach uncovers a $\ell^{-1}\log \ell$ that has yet to be proved. We gave an heuristic argument (discontinuities in the derivative of the ``regular'' part of the symbol) to explain its origin, but it would be interesting to perform a more rigorous and systematic study, especially in the more involved Toeplitz+Hankel case. In the CFT framework such occurrence requires some fine-tuning: it is a specific feature of corner free energies with angle $2\pi$, and shouldn't appear in correlations of local physical observables, or in the entanglement entropy. In general both approaches could be pushed much further, and we hope that our method can shed some more light on the general 
structure of subleading corrections to the Fisher-Hartwig formula, and its relation to the Renormalization Group in statistical mechanics. As an aside, we showed that the $\ell^{-1}\log \ell$ also appears in the full counting statistics for spins in the $z$ basis. 
 
 We also studied the XXZ chain, as an example of a model with conserved number of particles. Here the situation is somewhat incongruous, as the simplest correlation from the Algebraic Bethe ansatz point of view turns out to difficult to understand at the field-theoretical level. We have shown that usual boundary CFT arguments become insufficient, and performed extensive numerical computations in the free fermion case. We showed that the general scaling could be explained in terms of an arctic phenomenon, familiar in the study of dimers, and determined the frozen region numerically. We also provided some evidence for the universality of logarithmic corrections. A field-theoretical determination of these terms would require a better understanding of the fluctuations around the background we identified, and is left as an interesting open question. 

 Finally it might also be interesting to investigate similar setups for quantum quenches in real time. Indeed the conformal Ising case of Sec.~\ref{sec:cbc} bears some similarity to the Calabrese Cardy argument for global quenches \cite{CalabreseCardy2007}, where a system is prepared in the ground-state $\ket{\psi_0}$ of a massive Hamiltonian, but evolves at time $t>0$ with a critical Hamiltonian. The initial wave function $\ket{\psi_0}$ is then seen, in imaginary time, as sufficiently close \cite{Cardy_talk} to a conformal boundary state. Interestingly in the XXZ chain it is possible to study a slightly different problem where one starts from a strongly magnetized state (see e.g. Ref.~\cite{Antal,SabettaMisguich}), which is obviously very far from a conformal boundary state. This type of initial state is exactly the kind that generates an arctic phenomenon in imaginary time. It would therefore be interesting to investigate the possible influence of this arctic phenomenon in real time quench problems. 
 
 \ack
I wish to thank J\'erome Dubail for early collaboration on this project, and related work. I am also grateful to Karol Kozlowski for explaining me how his recent work can be generalized to treat the less regular Toeplitz symbols encountered in this study. Finally, I thank Alexander Abanov, Paul Fendley, Israel Klich, Gr\'egoire Misguich and Vincent Pasquier for stimulating discussions and suggestions. This work was supported by the U.S. National Science Foundation under the grant DMR/MPS1006549.

\clearpage
 \appendix
 \section[\;\;\;\;\;\;\;\;\;\;\;\;\;\;Numerical determination of the arctic region]{Numerical determination of the arctic region}
 \label{app:dimers}
 \subsection[\;\;\;\;\;\;\;\;\;\;\;\;\;\;The dimer problem]{The dimer problem}
\begin{figure}[htbp]
        \centering
        \begin{subfigure}[b]{0.44\textwidth}
                \centering
\includegraphics{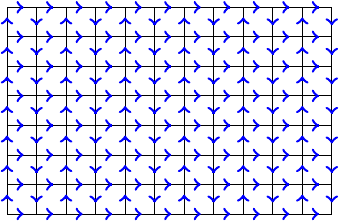}
                \caption{Kasteleyn orientation of the lattice: $K_{ij}=1$ if $(i\rightarrow j)$, $K_{ij}=-1$ if $(i\leftarrow j)$, $K_{ij}=0$ otherwise.}
                \label{fig:kasta}
        \end{subfigure}%
        \hfill
        \begin{subfigure}[b]{0.44\textwidth}
                \centering
\includegraphics{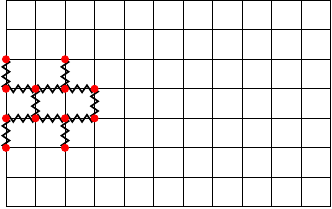}
                \caption{Links removed to ensure the $1,0,1,0$ dimer occupancies along an horizontal segment.}
                \label{fig:kastb}
        \end{subfigure}
        \caption{(a) Kasteleyn weighting of the lattice. (b) Probability of formation of a ``ferroelectric string'' of length $\ell=4$ with dimer occupancies $1,0,1,0$. The zigzag lines corresponds to the link we have to remove to ensure the ferroelectric configuration. The modified Kasteleyn matrix is obtained by simply setting to zero the corresponding matrix elements. The sites part of a removed link, and are represented in red circles. There are $n=3\ell$ of them. This will be the set ${\bf a}$ in Sec.~\ref{sec:perturbation}.}
        \label{fig:kast}
\end{figure}

 Let us consider the dimer problem on a square lattice with open boundary conditions. As is well known since Kasteleyn \cite{Kasteleyn}, the partition function (number of dimer converings) is given by
 \begin{equation}
  Z={\rm Pf}\, K =\sqrt{|\det K|}
 \end{equation}
where $K$ is a signed adjacency matrix. The matrix elements are shown in Fig.~\ref{fig:kasta}. For a given size $L_x\times L_y$, the Kasteleyn matrix can be diagonalized exactly, and its determinant evaluated in closed form. The local statistics can be handily accessed using $K$. For example, the probability to observe a given dimer on the link $(i,j)$ is given, up to a global sign, by
\begin{equation}
 p_{(i,j)}=K_{ij}\times (K^{-1})_{ij}
\end{equation}
$K$ can be inverted for finite $L_x,L_y$, we refer to \cite{FisherStephenson} for the exact expressions. In the limit $L_x,L_y\to\infty$ these expressions reduce to integrals that can be computed exactly (see again \cite{FisherStephenson}).  
We are interested in a system where an alternating sequence $1,0,1,0,\ldots $ of dimer occupancies along a segment is imposed somewhere on the left of the system, as is shown in Fig.~\ref{fig:kastb}. The number of dimer coverings compatible with such a condition is given by
\begin{equation}
 \tilde{Z}={\rm Pf}\,\tilde{K},
\end{equation}
where $\tilde{K}$ is deduced from $K$ by setting to zero the matrix elements corresponding to the removed links in Fig.~\ref{fig:kastb}. Inverting $\tilde{K}$ is more difficult, and accessing the $\tilde{p}_{(i,j)}$ should be much harder. There is however a simple trick, explained in the next section, that allows to overcome this difficulty.
 \subsection[\;\;\;\;\;\;\;\;\;\;\;\;\;\;Perturbed Kasteleyn matrices]{Perturbed Kasteleyn matrices}
 \label{sec:perturbation}
 $\tilde{K}$ is given by
 \begin{equation}
  \tilde{K}=K+E,
 \end{equation}
where $E$ is a ``perturbation'' matrix that has very few non zero matrix elements: $E_{ij}=-K_{ij}$ if the link $(i,j)$ has been removed, $0$ otherwise. The inverse may be written as
\begin{eqnarray}
 \tilde{K}^{-1}&=&\left[K+E\right]^{-1}\\
 &=&\left[K\left(1+K^{-1}E\right)\right]^{-1}\\
 &=&\left(1+K^{-1}E\right)^{-1}K^{-1}
\end{eqnarray}
Since we know how to calculate $K^{-1}$, let us study
\begin{equation}
 J=K^{-1}\tilde{K}=1+K^{-1}E
\end{equation}
in more details. Denote by ${\bf a}$ the set of sites part of a removed link. There are $n=3\ell$ of them. The other sites (the majority) are in ${\bf b}$, and there are $N-n$ of them. The matrix $J$ is close to the identity, in the sense that it differs from it only for $n$ columns. Indeed, 
\begin{equation}
 (K^{-1}E)_{ij}=\sum_k K^{-1}_{ik}E_{kj}
\end{equation}
is non zero only if $j\in {\bf a}$. Now we write $J$ in block form as
\begin{equation}
 J=\left(\begin{array}{ccc}
    J_{\bf aa}&& 0_{\bf ab}\\&&\\
    J_{\bf ba}&&1_{\bf bb}
   \end{array}\right),
\end{equation}
where $0_{\bf ab}$ is a $n\times (N-n)$ with zero matrix elements, and $1_{\bf bb}$ is the $(N-n)\times (N-n)$ identity matrix. Using this form, the emptiness formation probability simplifies into
\begin{equation}
 \left(p_0\right)^2=\frac{\det \tilde{K}}{\det K}=\det (K^{-1}\tilde{K})=\det J=\det J_{\bf aa}.
\end{equation}
Such types of formula have already been derived for the monomer correlation function on the square \cite{FisherStephenson} and triangular \cite{FendleyMoessnerSondhi} lattice. Inverting $J$, we get
\begin{equation}
 J^{-1}=\left(\begin{array}{ccc}
    \left(J_{\bf aa}\right)^{-1}&& 0_{\bf ab}\\&&\\
    -J_{\bf ba}\left(J_{\bf aa}\right)^{-1}&&1_{\bf bb}
   \end{array}\right).
\end{equation}
Notice $(J_{\bf aa})^{-1}=\left(J^{-1}\right)_{\bf aa}$, and thus the notation $J_{\bf aa}^{-1}$ is unambiguous. This is a huge simplification, because numerically inverting $J$ actually only requires to invert a small block $J_{\bf aa}$, the rest following from matrix multiplication. Similarly if we write $K^{-1}$ in block form, we finally get
\begin{equation}
 \tilde{K}^{-1}=\left(\begin{array}{ccc}
    J_{\bf aa}^{-1}\left(K^{-1}\right)_{\bf aa}&&  J_{\bf aa}^{-1}\left(K^{-1}\right)_{\bf ab}\\&&\\
    \left(K^{-1}\right)_{\bf ba}-J_{\bf ba}J_{\bf aa}^{-1}\left(K^{-1}\right)_{\bf aa}&&\left(K^{-1}\right)_{\bf bb}-J_{\bf ba} J_{\bf aa}^{-1}\left(K^{-1}\right)_{\bf ab}
   \end{array}\right).
\end{equation}
The crucial piece of information is encoded in $J_{\bf aa}$. In the end, the probability of interest reads
\begin{equation}
 \tilde{p}_{(ij)}=p_{(i,j)}-K_{ij}\sum_{p\in \mathbf{a}}\sum_{q\in \mathbf{a}} J_{ip} \left(J_{\bf aa}^{-1}\right)_{pq}\left(K^{-1}\right)_{qj},
\end{equation}
e. g. for $i \in \mathbf{b}, j\in \mathbf{b}$. The numerical evaluation of such double sums can be done relatively fast.
By that we mean that all the probabilities can be obtained in a time of order $\mathcal{O}(\ell^2 L_x L_y)$, on a $L_x\times L_y$ grid. Let us finally mention that inverting $J_{\bf aa}$ is numerically very unstable. Typically, the system sizes shown in this paper require close to a hundred digits of precision.
\subsection[\;\;\;\;\;\;\;\;\;\;\;\;\;\;Exact expressions for the propagators]{Exact expressions for the propagators}
The exact expressions are, with $i=i_x+(i_y-1)L_x$ and $j=j_x+(j_y-1)L_x$:
\begin{eqnarray}\fl
 K^{-1}_{ij}&=&K^{-1}(i_x,i_y;j_x,j_y)\\\fl
 &=&\frac{4(-1)^{\frac{1}{2}(1+j_x-i_x+j_y-i_y)}}{(L_x+1)(L_y+1)}\sum_{k_x}\sum_{k_y}
 \frac{\sin \left( i_x k_x\right)\sin\left(j_x k_x\right)\sin\left(i_y k_y\right)\sin\left(j_y k_y\right)}
 {\cos^2\left(k_x\right)+\cos^2\left(k_y\right)}
 \times f(k_x,k_y)
\end{eqnarray}
with
\begin{equation}
 f(k_x,k_y)=\left\{\begin{array}{ccc}\cos\left(k_x\right)&;&(j_x-i_x)\;{\rm odd}, (j_y-i_y) \;{\rm even}\\&&\\(-1)^{i_x+1}\cos (k_y)&;&(j_x-i_x)\;{\rm even}, (j_y-i_y) \;{\rm odd}\\&&\\0&;&{\rm otherwise}\end{array}\right.
\end{equation}
and
\begin{eqnarray}
 k_x=\frac{m_x\pi}{L_x+1}&,&m_x=1,2,\ldots,L_x/2\\
 k_y=\frac{m_y\pi}{L_y+1}&,&m_y=1,2,\ldots,L_y.
\end{eqnarray}
In the limit $L_x,L_y\to \infty$ these expressions reduce to integrals that can be computed exactly \cite{FisherStephenson}.


\clearpage
 \section*{References}
\begin{thebibliography}{99}
\bibitem{Bethe}
Bethe H, \journdoi{Zeitschrift f\"ur Physik}{71}{205}{1931}{Zur Theorie der Metalle}{10.1007/BF01341708}
\bibitem{Gaudin}
Gaudin M, {\it La fonction d'onde de Bethe} 1983 ed. Masson, France 
\bibitem{Quantuminverse}
Korepin V. E, Bogoliubov N. M and Izergin A. G, {\it Quantum Inverse Scattering Mehod and Correlation Functions} 1993, Cambridge University Press, Cambridge, UK 
\bibitem{Jimbo1}
Davies B, Foda O, Jimbo M, Miwa T and Nakayashiki A, \journdoi{Comm. Math. Phys.}{151}{89}{1993}{Diagonalization of the XXZ Hamiltonian by Vertex Operators}{10.1007/BF02096750}
\bibitem{Jimbo2}
Jimbo M and Miwa T, \journlink{Amer. Math. Soc. Conf}{85}{1-152}{1994}{Algebraic Analysis of Solvable Lattice Models}{http://cds.cern.ch/record/263995}
\bibitem{Lyon1}
Kitanine N, Maillet J-M and Terras V, \journdoi{Nucl. Phys. B}{554}{647}{1999}{Form factors of the XXZ Heisenberg spin-1/2 finite chain}{10.1016/S0550-3213(99)00295-3}
\bibitem{Lyon2}
Kitanine N, Maillet J-M and Terras V, \journdoi{Nucl. Phys. B}{567}{554}{2000}{Correlation functions of the XXZ Heisenberg spin-1/2 chain in a magnetic field}{10.1016/S0550-3213(99)00619-7}
\bibitem{BPZ}
Belavin A. A, Polyakov A. M and Zamolodchikov A. B, \journdoi{Nucl. Phys. B}{241}{333}{1984}{Infinite conformal symmetry in two-dimensional quantum field theory}{10.1016/0550-3213(84)90052-X}
\bibitem{BIK}
Bogoliubov N. M, Izergin A. G and Korepin V. E, \journdoi{Nucl. Phys. B}{275}{687}{1986}{Critical exponents for integrable models}{10.1016/0550-3213(86)90579-1} 
\bibitem{Lyon3}
Lukyanov S and Terras V ,\journdoi{Nucl. Phys. B}{654}{323}{2003}{Long-distance asymptotics of spin–spin correlation functions for the XXZ spin chain}{10.1016/S0550-3213(02)01141-0}
\bibitem{Lyon4}
Kitanine N, Kozlowski K. K, Maillet J. M, Slavnov N. A and Terras, V, \journdoi{J. Stat. Mech.}{}{P04033}{2009}{Algebraic Bethe ansatz approach to the asymptotic behavior of correlation functions}{10.1088/1742-5468/2009/04/P04003}
\bibitem{EFP_first}
Korepin V. E, Izergin A. G, Essler F. H. L and Uglov D. B, \journdoi{Phys. Lett. A}{190}{182}{1994}{Correlation function of the spin-1/2 XXX antiferromagnet}{10.1016/0375-9601(94)90074-4}
\bibitem{EFP_first2}
Essler F. H. L, Frahm H, Izergin A. G and Korepin V. E, \journdoi{Comm. Math. Phys.}{174}{191}{1995}{Determinant representation for correlation functions of spin-1/2 XXX and XXZ Heisenberg magnets}{10.1007/BF02099470}
\bibitem{EFP_XX}
Shiroishi M, Takahashi M and Nishiyama Y, \journdoi{J. Phys. Soc. Jpn.}{70}{3535}{2001}{Emptiness Formation Probability for the One-Dimensional Isotropic XY Model}{10.1143/JPSJ.70.3535}
\bibitem{EFP_XXZrazumov}
Kitanine N, Maillet J-M, Slavnov N and Terras V, \journdoi{J. Phys. A: Math. Gen}{35}{L385}{2002}{Emptiness formation probability of the XXZ spin-1/2 Heisenberg chain at Delta=1/2}{10.1088/0305-4470/35/27/102}
\bibitem{EFP_XXZrazumov2}
Kitanine N, Maillet J-M, Slavnov N and Terras V, \journdoi{J. Phys. A: Math. Gen.}{35}{L753}{2002}{Large distance asymptotic behaviour of the emptiness formation probability of the XXZ spin-1/2 Heisenberg chain}{10.1088/0305-4470/35/49/102}
\bibitem{AbanovKorepin}
Abanov A. G and Korepin V. E, \journdoi{Nucl. Phys. B.}{647}{565}{2002}{On the probability of ferromagnetic strings in antiferromagnetic spin chains}{10.1016/S0550-3213(02)00899-4}
\bibitem{KorepinLukyanov}
Korepin V. E, Lukyanov S, Nishiyama Y and Shiroishi M, \journdoi{Phys. Lett. A}{312}{21}{2003}{Asymptotic behavior of the emptiness formation probability in the critical phase of XXZ spin chain}{10.1016/S0375-9601(03)00616-9}
\bibitem{EFP_Kozlowski}
Kozlowski K. K, \journdoi{J. Stat. Mech}{P02}{006}{2008}{On the emptiness formation probability of the open XXZ spin-1/2 chain}{10.1088/1742-5468/2008/02/P02006}
\bibitem{EFP_Cantini}
Cantini L, \journdoi{J. Phys. A: Math. Theor.}{45}{135207}{2012}{Finite size emptiness formation probability of the XXZ spin chain at $\Delta =-\frac{1}{2}$}{10.1088/1751-8113/45/13/135207}
\bibitem{Ising_exponent}
Kaufman B and Onsager L, \journ{pr}{76}{1244}{1949}{Crystal Statistics, III. Short-Range Order in a Binary Ising Lattice}
\bibitem{FisherHartwig}
Fisher M. E and Hartwig R. E, \journdoi{Adv. Chem. Phys.}{15}{333}{1968}{Toeplitz determinants, some applications, theorems and conjectures}{10.1002/9780470143605.ch18}
\bibitem{FranchiniAbanov}
Franchini F and Abanov A. G., \journdoi{J. Phys. A: Math. Gen.}{38}{5069}{2005}{Asymptotics of Toeplitz determinants and the emptiness formation probability for the XY spin chain}{10.1088/0305-4470/38/23/002}
\bibitem{Lebowitz}
Diel H, {\it Field-theoretic approach to critical behaviour at surfaces} vol X, ed. C. Domb and J. L. Lebowitz, Academic Press, New-York 1986.
\bibitem{SD_2013}
St\'ephan J-M and Dubail J, \journdoi{J. Stat. Mech.}{}{P09002}{2013}{Logarithmic correction to the free energy from sharp corners with angle $2\pi$}{10.1088/1742-5468/2013/09/P09002} 
\bibitem{Kozlowski_conjecture}
Kozlowski K. K, {\it Truncated Wiener-Hopf operators with Fisher Hartwig singularities} \href{http://arxiv.org/abs/0805.3902}{arXiv:0805.3902}
\bibitem{Cardy1989}
Cardy J, \journdoi{Nucl. Phys. B.}{324}{581}{1989}{Boundary conditions, fusion rules and the Verlinde formula}{10.1016/0550-3213(89)90521-X}
\bibitem{CardyPeschel}
Cardy J and Peschel I, \journdoi{Nucl. Phys. B}{300}{377}{1988}{Finite-size dependence of the free energy in two-dimensional critical systems}{10.1016/0550-3213(88)90604-9}
\bibitem{Stephan_Ising}
St\'ephan, J-M, Misguich G and Pasquier V, \journ{prb}{82}{125455}{2010}{R\'enyi entropy of a line in two-dimensional Ising models}
\bibitem{Zaletel}
Zaletel M. P, Badarson J. H and Moore J. E, \journ{prl}{107}{020402}{2011}{Logarithmic Terms in Entanglement Entropies of 2D Quantum Critical Points and Shannon Entropies of Spin Chains} 
\bibitem{CalabreseCardy2004}
Calabrese P and Cardy J, \journdoi{J. Stat. Mech.}{}{P06002}{2004}{Entanglement entropy and quantum field theory}{10.1088/1742-5468/2004/06/P06002}
\bibitem{LiebSchultzMattis}
Lieb E. Schultz T and Mattis D, \journdoi{Ann. of. Phys.}{16}{407}{1961}{Two soluble models of an antiferromagnetic chain}{10.1016/0003-4916(61)90115-4}
\bibitem{Rectangle1}
Bondesan R, Dubail J, Jacobsen J. L and Saleur, \journdoi{Nucl. Phys. B.}{862}{553}{2012}{Conformal boundary state for the rectangular geometry}{10.1016/j.nuclphysb.2012.04.021}
\bibitem{DubailReadRezayi}
Dubail J, Read N, Rezayi, \journ{prb}{86}{245310}{2012}{Edge-state inner products and real-space entanglement spectrum of trial quantum Hall states}
\bibitem{CalabreseCardy2007}
Calabrese P and Cardy J, \journ{prl}{96}{136801}{2006}{Time Dependence of Correlation Functions Following a Quantum Quench}
\bibitem{QPotts}
Wu F. Y, \journdoi{Rev. Mod. Phys.}{54}{235}{1982}{The Potts Model}{10.1103/RevModPhys.54.235}
\bibitem{Goldenchain}
Feiguin A, Trebst S, Ludwig A. W. W, Troyer M, Kitaev A, Wang Z. and Freedman M. H., \journ{prl}{98}{160409}{2007}{Interacting Anyons in Topological Quantum Liquids: The Golden Chain}
\bibitem{AndrewsBaxterForrester}
Andrews G. E, Baxter R. J and Forrester J. P, \journdoi{J. Stat. Phys}{35}{193}{1984}{Eight-vertex SOS model and generalized Rogers-Ramanujan-type identities}{10.1007/BF01014383}
\bibitem{Vincent1}
Pasquier V, \journdoi{Nucl. Phys. B}{285}{162}{1987}{Two-dimensional critical systems labelled by Dynkin diagrams}{10.1016/0550-3213(87)90332-4}
\bibitem{Vincent2}
Pasquier V, \journdoi{J. Phys. A: Math. Gen.}{20}{L1229}{1987}{Lattice derivation of modular invariant partition functions on the torus}{10.1088/0305-4470/20/18/003}
\bibitem{Toeplitz_r0}
Simon B, {\it Orthogonal Polynomials on the Unit Circle} 2005 Amer. Math. Soc. {\bf 54}, Part 1.
\bibitem{Toeplitz_r1}
Krasovsky I, \journdoi{Prog. Prob.}{64}{305}{2011}{Aspects of Toeplitz determinants}{10.1007/978-3-0346-0244-0_16}
\bibitem{Toeplitz_r2}
Deift P, Its A and Krasovsky I, {\it Toeplitz matrices and Toeplitz determinants under the impetus of the Ising model. Some history and some recent results}, \href{http://arxiv.org/abs/1207.4990}{arXiv:1207.4990}
\bibitem{Szegostrong}
Sz\"ego G. \journnolink{Comm. Sem. Math. Univ. Lund.}{14}{228}{1952}{On certain Hermitian forms associated with the Fourier series of a positive function}
\bibitem{FisherStephenson}
Fisher M. E and Stephenson J, \journdoi{Phys. Rev.}{132}{1411}{1963}{Statistical Mechanics of Dimers on a Plane Lattice. II. Dimer Correlations and Monomers}{10.1103/PhysRev.132.1411}
\bibitem{Abanov_stats1}
Abanov A. G, Ivanov, D. A and Qian, Y \journdoi{J. Phys. A: Math. Theor}{44}{485001}{2011}{Quantum fluctuations of one-dimensional free fermions and Fisher–Hartwig formula for Toeplitz determinants}{10.1088/1751-8113/44/48/485001} 
\bibitem{Abanov_stats2}
Ivanov D. A, Abanov A. G and Cheianov V. V, \journdoi{J. Phys. A: Math. Theor.}{46}{085003}{2013}{Counting free fermions on a line: a Fisher–Hartwig asymptotic expansion for the Toeplitz determinant in the double-scaling limit}{10.1088/1751-8113/46/8/085003}
\bibitem{Sinekernel1}
Kitanine N, Kozlowski K. K, Maillet J-M, Slavnov N. A and Terras V, \journdoi{Comm. Math. Phys.}{291}{691}{2009}{Riemann-Hilbert Approach to a Generalised Sine Kernel and Applications}{10.1007/s00220-009-0878-1}
\bibitem{Sinekernel2}
Kozlowski K. K, {\it Riemann--Hilbert approach to the time-dependent generalized sine kernel}, 2011 Adv. Theor. Math. Phys. {\bf 15} \href{http://projecteuclid.org/euclid.atmp/1355321970}{1655}
\bibitem{Korepin1}
Jin B-Q and Korepin V. E, \journdoi{J. Stat. Phys.}{116}{79}{2004}{Quantum spin chain, Toeplitz determinants and Fisher-Hartwig conjecture}{10.1023/B:JOSS.0000037230.37166.42} 
\bibitem{Korepin2}
Its A R, Jin B-Q and Korepin V. E, \journdoi{J. Phys. A.}{38}{2975}{2005}{Entanglement in XY spin chain}{10.1088/0305-4470/38/13/011}
\bibitem{Korepin3}
Franchini F, Its A R and Korepin V E, \journdoi{J. Phys. A.}{41}{025302}{2008}{Renyi entropy of the XY spin chain}{10.1088/1751-8113/41/2/025302}
\bibitem{BasorToeplitz}
Basor E. L, \journnolink{Indiana. Univ. Math. J.}{6}{975}{1979}{A localization theorem for Toeplitz determinants}
\bibitem{Bottcher}
B{\"o}ttcher A, \journnolink{Z. Anal. Anw}{1}{23}{1982}{Toeplitz determinants with piecewise continuous generating functions}
\bibitem{EhrhardtPhD}
Ehrhardt T, {\it Toeplitz determinants with several Fisher-Hartwig singularities} PhD Thesis, Chemnitz, Germany (1997). 
\bibitem{BasorEhrhardt}
Basor E. L. and Ehrhardt T, \journdoi{Math. Nachr.}{228}{5}{2001}{Asymptotic formulas for determinants of a sum of finite Toeplitz and Hankel matrices}{10.1002/1522-2616(200108)228:1<5::AID-MANA5>3.0.CO;2-E}
\bibitem{BasorTracy}
Basor E. L and Tracy C. A, \journdoi{Physica. A.}{177}{167}{1991}{The Fisher-Hartwig conjecture and generalizations}{10.1016/0378-4371(91)90149-7}
\bibitem{DeiftItsKrasovsky}
Deift P, Its A and Krasovky I, \journdoi{Ann. of Math.}{174}{1243}{2011}{Asymptotics of Toeplitz, Hankel, and Toeplitz+Hankel determinants with Fisher-Hartwig singularities}{10.4007/annals.2011.174.2.12}
\bibitem{Barnes}
Barnes E. W. \journnolink{Quarterly journ. Pure and Appl. Math.}{31}{264}{1900}{The theory of the G-function}
\bibitem{CalabreseEssler}
Calabrese P and Essler, F, \journdoi{J. Stat. Mech.}{}{P08029}{2010}{Universal corrections to scaling for block entanglement in spin-1/2 XX chains}{10.1088/1742-5468/2010/08/P08029}
\bibitem{FagottiCalabrese}
Fagotti M and Calabrese P, \journdoi{J. Stat. Mech.}{}{P01017}{2011}{Universal parity effects in the entanglement entropy of XX chains with open boundary conditions}{10.1088/1742-5468/2011/01/P01017}
\bibitem{Gutman}
Gutman D. B, Gefen Y and Mirlin A.D, \journdoi{J. Phys. A: Math. Theor}{44}{165003}{2011}{Non-equilibrium 1D many-body problems and asymptotic properties of Toeplitz determinants}{10.1088/1751-8113/44/16/165003}
\bibitem{StephanPhD}
St\'ephan J-M, {\it Intrication dans des syst\`emes quantiques \`a basse dimension}, 2011 \href{http://ipht.cea.fr/Docspht//search/article.php?IDA=9761}{PhD thesis} (in french).
\bibitem{DubailStephan_2011}
Dubail J and St\'ephan J-M, \journdoi{J. Stat. Mech.}{}{L03002}{2011}{Universal behavior of a bipartite fidelity at quantum criticality}{10.1088/1742-5468/2011/03/L03002}
\bibitem{LL}
Giamarchi T, {\it Quantum physics in one dimension} Oxford University Press, New-York (2004).
\bibitem{LiebTM}
Lieb E. H, \journdoi{J. Math. Phys.}{8}{2339}{1967}{Solution of the Dimer Problem by the Transfer Matrix Method}{10.1063/1.1705163}
\bibitem{Ikhlef}
Alet F, Ikhlef Y, Jacobsen J-L, Misguich G and Pasquier V, \journdoi{Phys. Rev. E}{74}{041124}{2006}{Classical dimers with aligning interactions on the square lattice}{10.1103/PhysRevE.74.041124}
\bibitem{Stephan09}
St\'ephan J-M, Furukawa S, Misguich G and Pasquier V, \journ{prb}{80}{184421}{2009}{Shannon and entanglement entropies of one- and two-dimensional critical wave functions} 
\bibitem{Widom_circular}
Widom H, \journdoi{Indiana. Univ. Math. J.}{21}{277}{1971}{The Strong Szeg\"o Limit Theorem for Circular Arcs}{10.1512/iumj.1971.21.21022}
\bibitem{Dyson}
Dyson F, \journdoi{J. Math. Phys.}{3}{157}{1962}{Statistical Theory of the Energy Levels of Complex Systems. II}{10.1063/1.1703774}
\bibitem{Kasteleyn}
Kasteleyn P. W, \journdoi{Physica}{27}{1209}{1961}{The statistics of dimers on a lattice: I. The number of dimer arrangements on a quadratic lattice}{10.1016/0031-8914(61)90063-5}
\bibitem{JPS}
Jokush W, Propp J and Shor P. {\it Random Domino Tilings and the Arctic Circle Theorem} \href{http://arxiv.org/abs/math/9801068}{arXiv:math/9801068}
\bibitem{Kenyon1}
Cohn H, Kenyon R and Propp J, {\it A variational principle for domino tillings} 2001 J. Amer. Phys. Soc. {\bf 14} \href{http://www.ams.org/journals/jams/2001-14-02/S0894-0347-00-00355-6/home.html}{297}  
\bibitem{Kenyon2}
Sheffield S, Kenyon R and Okounkov A, {\it Dimers and amoebae} 2006 Ann. Math. {\bf 163} \href{http://www.jstor.org/stable/10.2307/20159982}{1019}
\bibitem{dwbc1}
Izergin A. G Coker D. A and Korepin V. E, \journdoi{J. Phys. A: Math. Gen.}{25}{4315}{1992}{Determinant formula for the six-vertex model}{10.1088/0305-4470/25/16/010}
\bibitem{dwbc2}
Korepin V. E and Zinn-Justin P, \journdoi{J. Phys. A: Math. Gen}{33}{7053}{2000}{Thermodynamic limit of the six-vertex model with domain wall boundary conditions}{10.1088/0305-4470/33/40/304}
\bibitem{dwbc3}
Colomo F. Noferini V. and Pronko A. G. \journdoi{J. Phys. A: Math. Theor}{44}{195201}{2011}{Algebraic arctic curves in the domain-wall six-vertex model}{10.1088/1751-8113/44/19/195201} 
\bibitem{dwbc_as1}
Zinn-Justin P, \journdoi{Phys. Rev. E}{62}{3411}{2000}{Six-vertex model with domain wall boundary conditions and one-matrix model}{10.1103/PhysRevE.62.3411}
\bibitem{dwbc_as2}
Bleher P. M and Fokin V. V, \journdoi{Comm. Math. Phys.}{268}{223}{2006}{Exact Solution of the Six-Vertex Model with Domain Wall Boundary Conditions. Disordered Phase}{10.1007/s00220-006-0097-y}
\bibitem{Helfgott}
Helfgott H. {\it Edge Effects on Local Statistics in Lattice Dimers: A Study of the Aztec Diamond (Finite Case)}, \href{http://arxiv.org/abs/math/0007136}{arXiv:math/0007136}
\bibitem{J1J2}
Okamoto K and Nomura K, \journdoi{Phys. Lett. A}{169}{433}{1992}{Fluid-dimer critical point in S=1/2 antiferromagnetic Heisenberg chain with next nearest neighbor interactions}{10.1016/0375-9601(92)90823-5}
\bibitem{xzmagn}
Eisler V, R\'acz Z and van Wijland, F, \journdoi{Phys. Rev. E}{67}{056129}{2003}{Magnetization distribution in the transverse Ising chain with energy flux}{10.1103/PhysRevE.67.056129}
\bibitem{Gritsevetal}
Gritsev V, Altman E, Demler E and Polkovnikov A, \journdoi{Nat. Phys.}{2}{705}{2006}{Full quantum distribution of contrast in interference experiments between interacting one-dimensional Bose liquids}{doi:10.1038/nphys410}
\bibitem{zmagn}
Cherng, R. W and Demler E, \journdoi{New. J. Phys.}{9}{7}{2007}{Quantum noise analysis of spin systems realized with cold atoms}{doi:10.1088/1367-2630/9/1/007}
\bibitem{SokalSalas}
Salas J and Sokal A. D, \journdoi{J. Stat. Phys.}{98}{551}{2000}{Universal Amplitude Ratios in the Critical Two-Dimensional Ising Model on a Torus}{10.1023/A:1018611122166} 
\bibitem{xmagn}
Lamacraft A and Fendley P, \journ{prl}{100}{165706}{2008}{Order Parameter Statistics in the Critical Quantum Ising Chain}
\bibitem{AbanovIvanov}
Ivanov D. A and Abanov A. G \journdoi{Phys. Rev. E}{87}{022114}{2013}{Characterizing correlations with full counting statistics: Classical Ising and quantum XY spin chains}{10.1103/PhysRevE.87.022114}
\bibitem{KlichLevitov}
Klich I and Levitov L, \journ{prl}{102}{100502}{2009}{Quantum Noise as an Entanglement Meter}
\bibitem{KlichYale1}
Song H. F, Flindt C, Rachel S, Klich I and Le Hur K, \journ{prb}{83}{161408}{2011}{Entanglement from Charge Statistics: Exact Relations for Many-Body Systems}
\bibitem{KlichYale2}
Song H, F, Rachel S, Flindt C, Klich I, Laflorencie N and Le Hur K, \journ{prb}{85}{035409}{2012}{Bipartite Fluctuations as a Probe of Many-Body Entanglement}
\bibitem{Pfeuty}
Pfeuty P, \journdoi{Ann. Phys.}{57}{79}{1970}{The one-dimensional Ising model with a transverse field}{10.1016/0003-4916(70)90270-8}
\bibitem{Barouch}
Barouch E and Mc Coy B. M, \journ{pra}{3}{786}{1971}{Statistical Mechanics of the XY Model. II. Spin-Correlation Functions}
\bibitem{FendleyFisherNayak}
Fendley P, Fisher, M. P. A and Nayak, C., \journdoi{Ann. Phys.}{325}{1547}{2009}{Boundary conformal field theory and tunneling of edge quasiparticles in non-Abelian topological states}{10.1016/j.aop.2009.03.005}
\bibitem{OshikawaAffleck}
Oshikawa, M. and Affleck, I., \journdoi{Nucl. Phys. B.}{95}{533}{1997}{Boundary conformal field theory approach to the critical two-dimensional Ising model with a defect line}{10.1016/S0550-3213(97)00219-8}
\bibitem{ee_corrections_bulk}
Cardy J and Calabrese P, \journdoi{J. Stat. Mech.}{}{P04023}{2010}{Unusual corrections to scaling in entanglement entropy}{10.1088/1742-5468/2010/04/P04023}
\bibitem{ee_corrections_boundary}
Eriksson E and Johannesson H, \journdoi{J. Stat. Mech.}{}{P02008}{2011}{Corrections to scaling in entanglement entropy from boundary perturbations}{doi:10.1088/1742-5468/2011/02/P02008}
\bibitem{Stephan_phasetransition}
St\'ephan J-M, Misguich G and Pasquier V, \journ{prb}{84}{195128}{2011}{Phase transition in the R\'enyi-Shannon entropy of Luttinger liquids}
\bibitem{Cardybcc}
Cardy J, \journdoi{Nucl. Phys. B}{275}{200}{1986}{Effect of boundary conditions on the operator content of two-dimensional conformally invariant theories}{10.1016/0550-3213(86)90596-1}
\bibitem{Rectangle2}
Bondesan R, Jacobsen J. L and Saleur H, \journdoi{Nucl. Phys. B.}{867}{913}{2013}{Rectangular amplitudes, conformal blocks, and applications to loop models}{10.1016/j.nuclphysb.2012.10.018}
\bibitem{Stephan_rvb}
St\'ephan J-M, Ju H, Fendley P and Melko R. G, \journdoi{New. J. Phys.}{15}{015004}{2013}{Entanglement in gapless resonating-valence-bond states}{10.1088/1367-2630/15/1/015004}
\bibitem{Ferdinand}
Ferdinand A. E, \journdoi{J. Math. Phys.}{8}{2332}{1967}{Statistical Mechanics of Dimers on a Quadratic Lattice}{10.1063/1.1705162}
\bibitem{Hinrichsen}
Um J, Park J and Hinrichsen H, \journdoi{J. Stat. Mech.}{}{P10026}{2012}{Entanglement versus mutual information in quantum spin chains}{10.1088/1742-5468/2012/10/P10026}
\bibitem{Grassberger}
Lau H. W and Grassberger P, \journdoi{Phys. Rev. E}{87}{022128}{2013}{Information theoretic aspects of the two-dimensional Ising model}{
10.1103/PhysRevE.87.022128}
\bibitem{SB}
Saleur H and Bauer M, \journdoi{Nucl. Phys. B}{320}{591}{1989}{On some relations between local height probabilities and conformal invariance}{10.1016/0550-3213(89)90014-X}
\bibitem{AOS}
Affleck I, Oshikawa M and Saleur H, \journdoi{J. Phys. A: Math. Gen.}{31}{5827}{1998}{Boundary critical phenomena in the three-state Potts model}{10.1088/0305-4470/31/28/003}
\bibitem{blockT1}
Widom H, \journnolink{Adv. Math. }{13}{284}{1974}{Asymptotic behavior of block Toeplitz matrices and determinants}
\bibitem{blockT2}
Widom H, \journnolink{Proc. Amer. Soc. }{50}{167}{1975}{On the limit of block Toeplitz determinants}
\bibitem{blockT3}
Its A. R, Jin B-Q and Korepin V. E \journnolink{Fields. Inst. Comm.}{50}{151}{2007}{Entropy of $XY$ spin chain and Block Toeplitz Determinants}
\bibitem{Cardy_talk}
Cardy J, {\it Quantum quenches in Conformal Field Theory from a General Short-Range State} 2012, \href{http://www-thphys.physics.ox.ac.uk/people/JohnCardy/seminars/GGI.pdf}{Talk at GGI}, Florence
\bibitem{Antal}
Antal T, Racz Z, Rakos A and Sch\"utz, \journdoi{Phys. Rev. E}{59}{4912}{1999}{Transport in the XX chain at zero temperature: Emergence of flat magnetization profiles}{10.1103/PhysRevE.59.4912}
\bibitem{SabettaMisguich}
Sabetta T and Misguich G, \journ{prb}{88}{245114}{2013}{Nonequilibrium steady states in the quantum XXZ spin chain}
\bibitem{FendleyMoessnerSondhi}
Fendley P, Moessner R and Sondhi S L \journ{prb}{66}{214513}{2002}{Classical dimers on the triangular lattice}
\end{thebibliography}
\end{document}